\documentclass[twocolumn,floatfix]{aastex631}

\usepackage{amssymb}
\usepackage{amsmath}
\usepackage[caption=false]{subfig}
\graphicspath{{figures/}}
\usepackage{algorithm}
\usepackage[linesnumbered,ruled,lined,algo2e]{algorithm2e}

\usepackage{silence}
\begin{document}

\title{\texttt{APPLE}: An Evolution Code for Modeling Giant Planets}

\shorttitle{\texttt{APPLE}: A Modern Code for Giant Planet Evolution}
\shortauthors{Sur et al.}

\correspondingauthor{Ankan Sur}
\email{ankan.sur@princeton.edu}

\author[0000-0001-6635-5080]{Ankan Sur}
\author[0000-0001-8283-3425]{Yubo Su}
\author[0000-0001-6708-3427]{Roberto Tejada Arevalo}
\author[0000-0003-3792-2888]{Yi-Xian Chen}
\affiliation{Department of Astrophysical Sciences, Princeton University, 4 Ivy Lane,
Princeton, NJ 08544, USA}
\author[0000-0002-3099-5024]{Adam Burrows}
\affiliation{Department of Astrophysical Sciences, Princeton University, 4 Ivy Lane,
Princeton, NJ 08544, USA}
\affiliation{Institute for Advanced Study, 1 Einstein Drive, Princeton, NJ 08540}

\newcommand{\gcc}{g cm$^{-3}$}
\newcommand{\Dmlt}{\mathcal{D}_{\rm MLT}}
\newcommand{\flux}{\mathcal{F}}
\newcommand{\vdag}{(v)^\dagger}
\newcommand\aastex{AAS\TeX}
\newcommand\latex{La\TeX}
\newcommand{\logpar}[4]{\bigg(\frac{\partial \log{#1}}{\partial \log{#2}}\bigg)_{#3, #4}}
\newcommand{\logd}[2]{\frac{d\log{#1}}{d\log{#2}}}
\newcommand{\parderiv}[4]{\bigg(\frac{\partial #1}{\partial #2}\bigg)_{#3, #4}}
\newcommand{\fullderiv}[2]{\frac{d #1}{d #2}}
\newcommand{\kt}{\kappa_T}
\newcommand{\ku}{\kappa_{\mu}}
\newcommand{\Nu}{\mathrm{Nu}}
\newcommand{\R}{\rm{R}_0^{-1}}
\newcommand{\Rc}{\rm{R}_c^{-1}}

\newcommand{\apple}{\texttt{APPLE}}


\begin{abstract}
We introduce \texttt{APPLE}, a novel planetary evolution code designed specifically for the study of giant exoplanet and Jovian planet evolution in
the era of \textit{Galileo}, \textit{Juno}, and \textit{Cassini}. With \texttt{APPLE}, state-of-the-art equations of state for hydrogen, helium, ice, and rock are integrated with advanced features to treat ice/rock cores and metals in the gaseous envelope; models for helium rain and hydrogen/helium immiscibility; detailed atmosphere boundary tables that also provide self-consistent albedos and spectra; and options to address envelope metal gradients and stably-stratified regions. Our hope is that these purpose-built features of \texttt{APPLE} will help catalyze the development of the next generation of giant exoplanet and Jovian planet evolutionary models.
\end{abstract}

\section{Introduction} 
\label{sec:intro}

The first giant exoplanet evolutionary models were created by \cite{Burrows1995}. However, the characteristics of these models were informed by earlier evolutionary models of the Jovian planets of our solar system. Hence, they assumed adiabaticity, homogeneity, and simple atmospheres tied to interior adiabats \citep{Hubbard1968, Hubbard1969, Hubbard1977}. \citet{Burrows1997}, \citet{Baraffe1998}, \citet{Burrows2001}, and \citet{Baraffe2003} expanded upon this work with non-gray atmospheric models as boundary conditions. From the outset, giant exoplanet evolutionary and spectral models were conceived in the context of a wider giant planet/brown dwarf continuum extending from $\sim$0.1 to $\sim$80 $M_J$ (Jupiter mass).  A more modern realization of this tradition with updated molecular opacities and atmospheres is the ``Sonora-Bobcat" model suite of \citet{marley2021_sonora}.  However, to date, such models generally still assume adiabaticity, lack heavy-element cores, do not incorporate helium rain \citep{Stevenson1977a}, and assume homogeneity.

The early insights into Jovian planet properties were obtained by the $Voyager$ \citep{Hanel1981, Hanel1983, Campbell1985} and $Pioneer$ \citep{Ingersoll1975} probes.  These were followed by the \textit{Galileo} entry probe \citep{vonZahn1992, Niemann1992, Young1996}, whose results have significantly enhanced our understanding of giant planet formation and evolution. In particular, \textit{Galileo} provided estimates of Jupiter's atmospheric composition (generally $\sim$3$\times$solar), but also indicated a depletion in atmospheric helium (to a mass fraction, $Y$, of $\sim$0.234) \citep{vonZahn1998} compared to the protosolar value of $\sim0.272$\footnote{This is the helium mass fraction, $Y$. Often, the helium abundance is given as $\frac{Y}{(1-Z)}$, where $Z$ is the metallicity. Under this convention, which we do not follow, the initial value of $\frac{Y}{1-Z}$ is approximately 0.2777 \citep{Bahcall2006}.} \citep{Bahcall2006}, hinting at the possibility of helium rain, as proposed by \cite{Stevenson1977a, Stevenson1977b} in the context of Saturn's ``anomalous" luminosity. More recently, missions such as \textit{Juno} \citep{Bolton2017a, Bolton2017b} and \textit{Cassini} \citep{Spilker2019} have measured the gravitational moments of Jupiter and Saturn \citep{Iess2019, Durante2020}, leading to the inference of chemically inhomogeneous models supporting the idea of an extended ``fuzzy core" \citep{Wahl2017, Debras2019, Nettelmann2021, Militzer2022, Helled2024}. Moreover, \textit{Cassini}, using Saturn ring seismology, suggests the presence in Saturn itself of g-mode oscillations and, hence, of a partially stably-stratified interior and dilute core there as well \citep{Fuller2014, Mankovich2021}. These results for both Jupiter and Saturn highlight the limitations of adiabatic modeling in the presence of composition gradients \citep{Leconte2012, Vazan2016, Berado2017}, as efficient convection would otherwise homogenize composition. They also punctuate the importance of more carefully addressing helium rain and hydrogen/helium immiscibility, as well as compositional inhomogeneity and non-adiabatic interiors. 

Multiple research groups \citep{Fortney2003, Pustow2016, Mankovich2016, Mankovich2020} have recently addressed the challenge of hydrogen/helium immiscibility in Jupiter and Saturn by employing a variety of miscibility curves \citep[e.g.][]{Hubbard1985, Pfaffenzeller1995, Lorenzen2011, Schottler2018} and redistribution schemes. Most of these models were ad-hoc, some implemented by hand, and a more self-consistent treatment of helium rain coupled with modern immiscibility curves is now called for. Moreover, compositional inhomogeneity and non-adiabaticity may lead to stably-stratified layers in which doubly-diffusive (DD) (semi-)convection may be triggered \citep{Rosenblum2011, Mirouh2012, Leconte2012, Wood2013, Nettelmann2015}. Energy transfer and compositional mixing within these layers would then be partially suppressed compared to overturning (``Ledoux") convection \citep{Chabrier1997,Leconte2012, Leconte2013}, with implications for the evolution of a planet's luminosity and effective temperature ($T_{\rm eff}$). How to properly handle such physics in an evolutionary code has yet to be clearly determined \citep{Vazan2015, Vazan2016}. \cite{Nettelmann2015} was the first to link Jupiter's thermal evolution with helium rain and DD convection while constraining the sizes of semi-convective layers to match its heat flux. This work has more recently been followed by \cite{Mankovich2016}; \citet{Moll2016}; \citet{Mankovich2020}; and \citet{Fuentes2022}. However, much work remains to be done to capture this effect with fidelity in an evolutionary context.

Furthermore, giant planet evolutionary calculations require careful consideration of atmospheric boundary conditions. Previous methodologies encompassed various approaches, including the non-gray isolated-object grid of model atmospheres by \cite{Burrows1997}, the simpler model atmosphere grid proposed by \cite{Graboske1975}, and the radiative-atmosphere grid presented by \cite{Fortney2011}. Calculations often assume a Bond albedo and employ the approximate equation $T_{\rm eff}^4 = T_{\rm int}^4 + T_{\rm eq}^4$ \citep[see e.g.][]{Baraffe2003, Fortney2003, Fortney2004, Mankovich2016, Mankovich2020}. Earlier models also overlooked the influence of cloud condensates in their atmospheres, which could significantly affect cooling. To address this, self-consistent model atmospheres, such as those developed by \cite{Chen2023}, are needed. These models incorporate atmospheric radiative transfer, accommodate clouds of different particle sizes, and redistribute heat across day and night regions. They also implicitly provide geometric and Bond albedos, as well as emission spectra over a broad wavelength range. 

Finally, another crucial facet of exoplanet and Jovian planet evolutionary modeling is the Equation of State (EOS) for the hydrogen and helium mixture. In the past, the SCvH95 EOS \citep{Scvh1995} was used extensively in giant planet, giant exoplanet, and brown dwarf evolutionary codes \citep{Burrows1997, Fortney2003, Nettelmann2015, Mankovich2016}. The default setting in the \texttt{MESA} \citep{Paxton2011} stellar evolution code allows for the utilization of the SCvH95 EOS. However, with advancements in computational power and more sophisticated material physics techniques, a new generation of EOSes derived from ab initio simulations has become available \citep{Nettelmann2008, Militzer2008, Nettelmann2012, Militzer2013, Becker2014, Chabrier2019, Chabrier2021}. Although recent studies such as those of \cite{Pustow2016} and \cite{Mankovich2020} have adopted these EOSes, modern evolutionary tools should provide various EOS options within a unified framework.

Hence, recent discoveries by solar-system probes and improvements in the microphysics relevant to giant planet interiors highlight the need for new innovative approaches to giant planet evolutionary modeling. In addition, while current stellar evolution codes are sophisticated, they still face challenges in accurately simulating phenomena such as helium rain and in accommodating diverse EOSes and rocky cores. Furthermore, atmospheric physics for such codes, and for most codes used in the past to model giant planet evolution, remains rather limited. With these imperatives in mind, a new generation of evolution code is called for. In this spirit, we present here \texttt{APPLE}\footnote{which stands for {A Python code for PLanetary Evolution}}. \texttt{APPLE} is a uniform modeling platform for giant exoplanet and Jovian planet evolutionary simulations. In this paper, we explain its architecture, the numerical techniques employed, and the physical principles that undergird it. The versatility of \texttt{APPLE} is showcased through a series of test cases, illustrating its application across a range of planetary conditions and scenarios. The structure of \texttt{APPLE} is modular and its computational structure easily enables upgrades as new physics and techniques appear. 

This paper is organized as follows. \S\ref{sec:structure} provides the structure and evolution equations, while \S \ref{sec:boundary_conditions} discusses the associated boundary conditions. \S\ref{sec:transport} explains the physics of energy transport in planetary interiors and \S\ref{sec:mixing} discusses the mixing and diffusion of elements. We continue a discussion of the relevant microphysics in \S\ref{sec:microphysics} and in \S\ref{sec:atmospheres} we describe the model atmosphere approach for giant planets we employ. This methodology also provides albedos and emission spectra \citep{Chen2023}. We then go on in \S\ref{sec:methods} to discuss the computation methods used in \texttt{APPLE} and in \S \ref{sec:simulations} present example simulations. The full numerical scheme, including the finite difference and discretization procedures, is presented in the Appendix. A list of symbols with their definitions is presented in Table \ref{table:1}. We wrap things up in \S \ref{sec:summary} with a summary discussion and thoughts on future directions.   

\section{Structure and Evolution equations}
\label{sec:structure}

The equations that describe the evolution of planets parallel those employed in stellar evolution \citep{Eggleton1971, Demarque2007, Dalsgaard2008, Weiss2008, Morel2008, Dalsgaard2008, Paxton2011, Paxton2013, Paxton2018, Paxton2019} which are the equations of mass, momentum, energy, and species conservation. Dropping the acceleration (inertia) term in the momentum equation, assuming that transport of energy and species are diffusive processes, and assuming spherical symmetry, one arrives at a standard set of coupled differential equations:

\begin{align}
\frac{dP}{d M_r} &= -\frac{G M_r}{4\pi r^4} + \frac{\Omega^2}{6\pi r}
\label{eq:1}\\
\frac{dr}{d M_r} &= \frac{1}{4\pi r^2 \rho}
\label{eq:2}\\
\frac{\partial L}{\partial M_r} &=  -\frac{dU}{dt} - P\frac{d(\frac{1}{\rho})}{dt}
\label{eq:3a}\\
&= -T\frac{dS}{dt} - \sum_i \left(\frac{\partial U}{\partial X_i}\right)_{s,\rho}\frac{dX_i}{dt}
\label{eq:3b}\\
N_A\frac{dX_i}{dt} &= -\frac{\partial }{\partial M_r}\left(4\pi r^2 F_i\right)\, ,
%\rho  \mathcal{D}\frac{\partial Y}{\partial r}\right) + \frac{\partial F_D}{\partial M_r}
\label{eq:4}
%& F_D = 
%\begin{cases}
%4 \pi r^2\rho \mathcal{D} \frac{Y}{H_r} & \text{rain model: A}\\
%4 \pi r^2\rho \mathcal{D} \frac{{\rm max}(0, Y-Y_{\rm misc})}{H_r} &\text{rain model: B}
%\end{cases}
\end{align}
where $M_r$ is the independent Lagrangian coordinate denoting the mass enclosed in a sphere of radius $r$, $P$ is the pressure, $L = 4\pi r^2 \cal{F}$ is the luminosity, $\cal{F}$ is the energy flux, $S$ is the specific entropy, $\rho$ is the mass density, and $T$ is the temperature, $U$ is the specific internal energy, $N_A$ is Avogardo's number, $X_i$ is the mass fraction of species $i$, and $\Omega$ is the solid-body angular frequency\footnote{ We emphasize that the term we employ in the equation of hydrostatic equilibrium to capture the centrifugal effect of rotation is rather crude, though similar to what is employed in standard stellar evolution codes.  Going beyond this approach is beyond the scope of this study. Ideally, an evolutionary model for rotating giant planets
such as Jupiter and Saturn would include multi-D rotational effects.  However, there currently exists no planet code that addresses the evolution
of the thermal, structural, and atmospheric evolution of giant planets in
full 3D (or even 2D), and there is as yet no tradition for such codes on which to build (however, see \citet{Hubbard1970}). We note that the numerical methods that address the Juno, etc. gravitational moment data are static structural models without thermal evolution that do not address the co-evolution of the thermal and physical structures with the atmospheres.
Time-dependent, evolutionary models have perforce been limited
to spherical symmetry due in part to the complexity of rotational
evolution. Going beyond this would involve the incorporation of
angular-momentum transport due to viscous, turbulent,
and magnetic torques in a multi-dimensional context that is beyond the scope of this effort.  Moreover, energy and composition transport beyond the mixing-length prescription of stellar evolution might be needed. Whatever the generalization beyond the standard stellar evolution paradigms that we have followed to craft \texttt{APPLE}, the result would unlikely after $\sim$4.56 billion years land on a suite of current J$_i$ values that could compete with the precision of the measured Juno values. We note that
even the static models that have been employed to reproduce the Juno J$_i$ values (e.g., \citet{Wahl2017} and \citet{Militzer2024}) have a range
of inferred potential compositional and thermal profiles. Moreover, the thermal profiles are very poorly constrained in those inversions and the
results are not one-to-one. With \texttt{APPLE}, we have built
upon a decades-long tradition of codes addressing the thermal evolution of
giant planets, which includes the work of \citet{Hubbard1999}; \citet{Saumon1996}; \citet{Burrows1997}, \citet{Burrows2001}; \citet{marley2021_sonora}; \citet{Baraffe1998}; \citet{Fortney2011}; \citet{Nettelmann2015}; \citet{Pustow2016}; and \citet{Mankovich2016}, \citet{Mankovich2020}. None of these authors when doing evolutionary calculations have attempted in a consistent fashion to capture the rotational evolution of the gravitational moment data obtained from \textit{Galileo}, \textit{Juno}, and \textit{Cassini}.}. 
Note that these equations involve the specific quantities (e.g., per gram), not the per volume quantities. To approximately incorporate the possible angle-averaged structural effects of solid-body rotation (with angular frequency $\Omega$), we follow \citet{Kippenhahn2013} in introducing an additional term in eq. \ref{eq:1}. For our purposes, $X_i$ = \{$X,Y,Z$\}, where $X$, $Y$, and $Z$ are the mass fractions of hydrogen, helium, and heavy elements, respectively, and $\sum X_i = 1$. The formalism can easily be extended to follow multiple heavy species. $F_i$ is the flux of species $i$ (and has both diffusive and advective contributions) and will be discussed in the context of helium rain in \S\ref{sec:mixing} and \S\ref{subsec:helium_rain}. We note that, as is standard practice in spherical stellar and planetary evolution, we treat mixing processes using an equation with a diffusive form. This allows one to conserve the mass in a species to machine precision by differencing it using Gauss's theorem. A similar approach is followed for $Z$ mixing, but for brevity is not highlighted in this paper.  

We note that $\frac{\partial U}{\partial X_i}$ is directly related to the chemical potential ($\mu_i$), modulo Avogadro's number, and that there are two ways of writing the energy conservation equations eq. \ref{eq:3a} and eq. \ref{eq:3b}.
Equation \ref{eq:3a} is the standard statement of energy conservation that explicitly focuses on the change in internal energy (temperature) and the work done (``$PdV$"). The other focuses explicitly on the entropy change and composition changes. For adiabatic and fully convective behaviors, the latter is more natural and represents the early history of the field when giant planets were thought to be fully convective and the second term was finessed \citep{Fortney2003, Mankovich2020}. However, the \textit{Juno} and \textit{Cassini} data now strongly suggest that such pioneering models have seen their day and that composition gradients and changes in $X_i$ with time (such as accompanying helium rain and heavy element redistribution) must be incorporated. In \texttt{APPLE}, we employ the approach summarized with eq. \ref{eq:3b}.  We were motivated to use entropy, composition, and density/radius as our primary variables in our Lagrangian code by the continuing importance of convection and convective energy and species transport and map out our computational methodology in \S\ref{sec:methods}.

Since we focus on Jovian planets, which are primarily composed of hydrogen and helium, along with some fraction of heavy elements distributed in the core and the envelope, we must use radiative opacities and conductivities (described in \S\ref{subsec:rad_cond}), an algorithm for handling convective energy transport (described in \S\ref{sec:conv}), an approach to species mixing (described in \S\ref{sec:mixing}), hydrogen-helium and metal equations of state (described in \S\ref{subsec:eos}), and, importantly, a methodology for incorporating the atmosphere boundary conditions that drive the global evolution of the entire system (described in \S\ref{sec:atmospheres}).  \texttt{APPLE} has been constructed with multiple embedded sets of microphysical inputs and has been made modular and flexible to accommodate upgrades in the input physics as they arise. 

\begin{table}[!htbp]
\centering
\begin{tabular}{ l l } 
\hline\hline
Symbol & \hspace{1cm} Meaning \hspace{3cm} \\ [0.2ex] 
 \hline\hline
  $a$ & \hspace{1cm} Radiation constant\\ 
 %$\alpha$ & \hspace{1cm} Mixing length parameter \\ 
 $c$ & \hspace{1cm} Speed of light \\[0.1ex]
 $C_p$ & \hspace{1cm} Specific heat at constant pressure \\
 $C_v$ & \hspace{1cm} Specific heat at constant volume \\
 $\mathcal{D}_{\rm MLT}$ & \hspace{1cm} Diffusion coefficient set by MLT\\
 $\mathcal{D}_{\rm micro}$ & \hspace{1cm} Microscopic diffusion coefficients \\
  $\mathcal{D}_{\rm sc}$ & \hspace{1cm} Semiconvective diffusion coefficients \\
 $\mathcal{F}$ & \hspace{1cm} Energy flux\\
 $g$ & \hspace{1cm} Acceleration due to gravity \\ 
 $G$ & \hspace{1cm} Universal Gravitation constant \\ 
 $H_p$ & \hspace{1cm} Pressure scale height\\ 
 $H_r,\mathcal{H}_r$ & \hspace{1cm} Helium rain scale parameters\\ 
 $k_B$ & \hspace{1cm} Boltzmann constant \\ 
 $\kappa_R$ & \hspace{1cm} Rosseland mean opacity \\ 
 $L$ & \hspace{1cm} Luminosity \\ 
 $\lambda_{\rm cd}$ & \hspace{1cm} Thermal conductivity due to conduction \\ [0.1ex]
$\lambda_r$ & \hspace{1cm} Thermal conductivity due to radiation \\ [0.1ex]
 $M_r$ & \hspace{1cm} Mass enclosed at radius $r$  \\
 $M_p$ & \hspace{1cm} Mass of the planet  \\
 %${N_i}$ & \hspace{1cm} Number of particles of species i\\
 Nu$_{T}$ & \hspace{1cm} Thermal Nusselt number \\
 Nu$_{X}$ & \hspace{1cm} Compositional Nusselt number \\
 ${\nabla}$ & \hspace{1cm} $\equiv \frac{d\ln T}{d \ln P}$, Temperature gradient\\[0.1ex]
 ${\nabla_{\rm ad}}$ & \hspace{1cm} $\equiv \left.\frac{\partial \rm ln T}{\partial \rm ln P} \right|_{S}$, Adiabatic temperature gradient\\
 $\Omega$ & \hspace{1cm} Angular frequency \\ 
 $P$ & \hspace{1cm} Pressure \\  
 Pr & \hspace{1cm} Prandlt number\\  
 $\rho$ & \hspace{1cm} Density \\ 
 $r$ & \hspace{1cm} Radial coordinate \\ 
 $R_{\rho}, R_0$ & \hspace{1cm} Density ratio in semiconvection \\ 
 Ra$_{\rm T}$ & \hspace{1cm}  Rayleigh number\\ 
 $R_p$ & \hspace{1cm} Radius of the planet \\ 
 $\sigma$ & \hspace{1cm} Stefan-Boltzmann constant \\ 
 ${S}$ & \hspace{1cm} Specific entropy\\
 $T$ & \hspace{1cm} Temperature \\ 
 $T_{\rm int}$ & \hspace{1cm} Internal/Intrinsic temperature \\ 
 $T_{\rm eff}$ & \hspace{1cm} Effective temperature \\ 
 $\tau$ & \hspace{1cm} Lewis number \\ 
 ${U}$ & \hspace{1cm} Specific internal energy\\
 ${\mu_i}$ & \hspace{1cm} Chemical potential of species i\\
 $v_{\rm MLT}$ & \hspace{1cm} Velocity of a fluid parcel from MLT\\
 ${\chi_T}$ & \hspace{1cm} $\equiv -\frac{\partial \ln P}{\partial \ln T}\bigg|_{\rho,Y}$\\
 ${\chi_Y}$ & \hspace{1cm} $\equiv \frac{\partial \ln P}{\partial Y}\bigg|_{\rho,T}$ \\
  ${X}$ & \hspace{1cm} Mass fraction of hydrogen\\
 ${Y}$ & \hspace{1cm} Mass fraction of helium\\
 ${Y_{\rm atm}}$ & \hspace{1cm} Atmospheric helium abundance\\
 ${Y_{\rm misc}}$ & \hspace{1cm} Immiscibility Helium fraction\\
 ${Z}$ & \hspace{1cm} Mass fraction of heavy elements\\
 \hline
\end{tabular}
\caption{Some useful symbols used in this paper and the \texttt{APPLE} code.}
\label{table:1}
\end{table}

\section{Boundary conditions}
The two hydrostatic equilibrium equations \ref{eq:1} and \ref{eq:2} need two boundary conditions to be solved, namely, one for pressure and the other for the radius. Ideally, the outer pressure should be zero, and we set it to a very small number (for numerical convenience) compared to the interior pressures. Therefore, we have
\begin{align}
    & P_{\rm surf} (M_r = M_p) = ``1 \rm\ bar" \\
    & r_{\rm center} (M_r = 0) = 0\, ,
\end{align}
where $M_p$ is the total mass of the planet. To solve the evolution equations \ref{eq:3a} and \ref{eq:3b}, four boundary conditions must be met. For energy transport, at the outer boundary, the surface flux is given by
\begin{equation}
    \mathcal{F}_{\rm surf} = {\sigma T_{\rm int}^4}\, ,   
    \label{eqn:definition_Tint}
\end{equation}
where the internal temperature, ${T_{\rm int}}$ { (equal to the effective temperature, $T_{\rm eff}$, when the planet is not irradiated), is calculated using a sophisticated atmospheric model as a function of gravity and entropy that is fully described in \S\ref{sec:atmospheres}.} $R_p$ is the planet's radius at a given age. At the inner boundary, we set
\begin{equation}
    \mathcal{F}_{\rm center} (M_r=0) = 0\, .  
\end{equation}
Both at the core-envelope boundary and at the surface, we use zero flux boundary conditions for the diffusion equation \ref{eq:4}. This is achieved by setting the diffusion coefficient at the boundaries equal to  
\begin{equation}
    4\pi r^2 \rho \mathcal{D}\frac{\partial Y}{\partial r}\bigg|_{M_r=M_c} = \, 4\pi r^2 \rho \mathcal{D}\frac{\partial Y}{\partial r}\bigg|_{M_r=M_p} = 0\, ,
\end{equation}
where $M_c$ is the mass coordinate at the envelope-core boundary, interior to which resides the ``ice/rock" core (of flexible composition), and $\mathcal{D}$ is the relevant diffusivity parameter. If there is no core, $M_c = 0$.

\begin{figure*}
    \centering
    \includegraphics[scale=0.58]{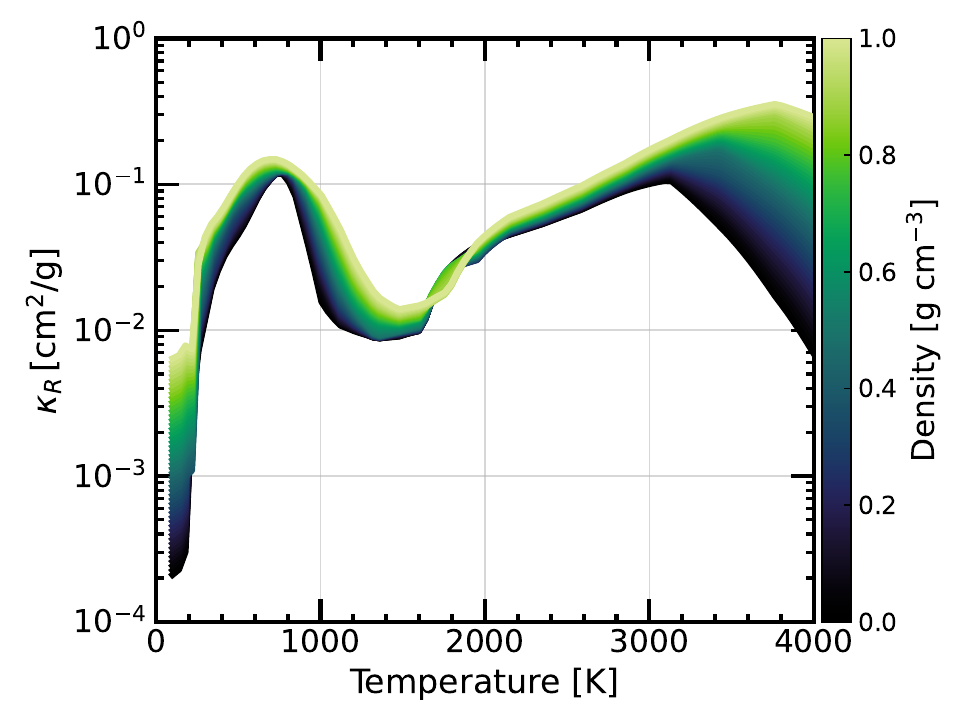}
    \includegraphics[scale=0.58]{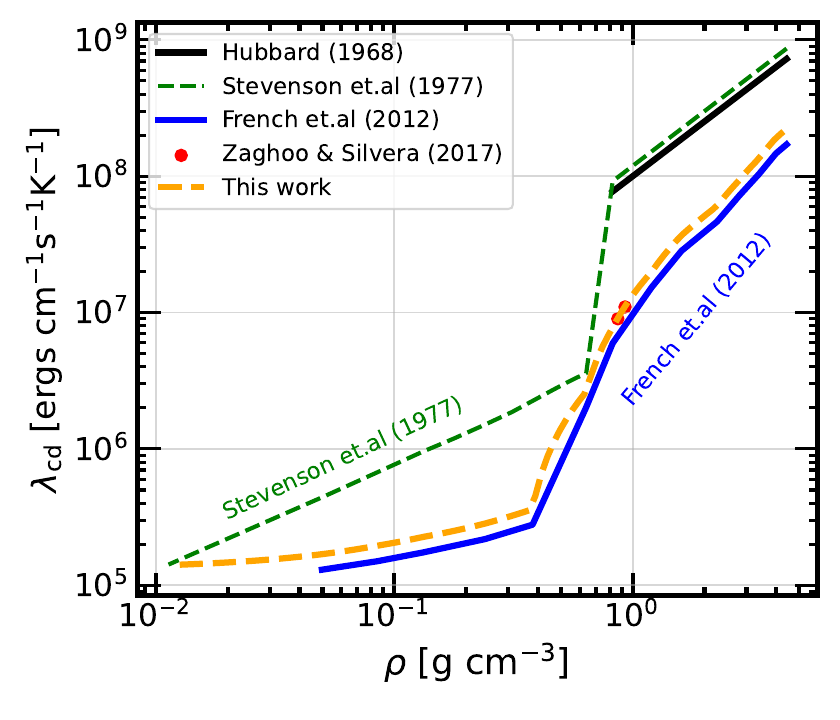}
    \caption{Left panel: Variation of Rosseland Mean Opacities ($\kappa_R$) with temperature and density. Right panel: Thermal conductivity $\lambda_{\rm cd}$ as a function of {density}. {The solid black line illustrates findings from \cite{Hubbard1968} in the metallic H region, with extensions into the molecular H region by \cite{Stevenson1977a} (green dashed), while the blue line depicts results from \cite{French2012} which have {superceded} the previous results. These results assume Y=0.275 and its corresponding helium number fraction of 0.0866. The \cite{Stevenson1977a} results are close to the \cite{Hubbard1968} results in the metallic regime, except for a factor 1.2 higher due to the assumed helium fraction.} The red dots are experimental values obtained by \citet{Zaghoo2017}. {We employed for this plot the Jupiter adiabatic profile provided in Table 1 of \citet{French2012} and interpolated the thermal conductivity in density. Multiplying this curve by 1.3 to match the \citet{Zaghoo2017} experimental values found at the two specific pressures explored in that work, the orange-dashed line represents the default currently in \texttt{APPLE}}.}
    \label{fig:cond_fig}
\end{figure*}

\label{sec:boundary_conditions}

\section{Energy transport}
\label{sec:transport}
\subsection{Radiation and Conduction}
\label{subsec:rad_cond}

A planet's structure and evolution are profoundly influenced by the mechanisms of heat transfer within their interiors and atmospheres \citep{Stevenson1982, Burrows1997, Marley1999, Hubbard1999, Burrows2001,Fortney2003, Fortney2011, Mankovich2016}. Conduction and radiation are basic processes that can play a key, though often subdominant, role.

Radiative transfer is an important mode of energy transport in stars and in the outer layers of planets \citep{Hubbard1968, Guillot1994a, Guillot1994b, Guillot1995, Chabrier1997, Arras2006, Robinson2012}. In the context of stellar and planetary evolution, radiation is treated as a diffusive process. The radiative flux is given by \citet{Kippenhahn1990}:
 \begin{equation}
     \flux_r = -\frac{4ac}{3}\frac{T^3}{\kappa_r \rho}\frac{\partial T}{\partial r} = -\lambda_r \frac{\partial T}{\partial r}\, , 
     \label{eq:rad_flux}
 \end{equation}
where $\kappa_r$ is the Rosseland mean coefficient and $\lambda_r = \frac{4ac}{3}\frac{T^3}{\kappa_r \rho}$ is the associated ``thermal conductivity". In \apple, we adhere to a conventional approach wherein we delineate the absorption characteristics and bound-bound transitions of atoms and molecules through pre-calculated, fixed tables containing Rosseland mean opacities, denoted as $\kappa_R(\rho, T)$ from \cite{Sharp2007} and updated recently by \citet{Lacy2023}, but the code can accommodate any opacity tables. The table spans a temperature range between $100\leq T\leq4000$ K while the density range is $10^{-5}\leq\rho\leq10^{-2}$ \gcc. The left panel of Figure \ref{fig:cond_fig} shows $\kappa_R$ as a function of $\rho$ and $T$. We note that it is only in the atmosphere that such transport is generally relevant to planetary evolution and that we already precalculate and tabulate a set of atmosphere boundary conditions to determine the total energy flux out of the planet and the outer flux boundary condition (see \S\ref{sec:atmospheres} and \S\ref{sec:boundary_conditions}). Nevertheless, including a radiative flux capability in the numerics is not double counting and allows us to incorporate radiative transfer effects in the inner regions, should they become of importance.  

Another mechanism of heat transfer in planetary interiors is conduction. This mode of energy transport is particularly important inside a planet's ``rocky" core \citep[see][for a detailed discussion]{Labrosse2015}, is generally more important than radiative transfer (except in the atmosphere), but is far less effective in transporting heat than convection (\S\ref{sec:conv}) or semiconvection (\S\ref{sub:semi}). The rate of conductive heat transfer is dictated by the temperature gradient and the material properties. The heat flux due to conduction is given by:

\begin{align}
    \flux_{\rm cd} 
    %&= -\frac{4ac}{3}\frac{T^3}{\kappa_{\rm cd} \rho}\frac{\partial T}{\partial r}
    = -\lambda_{\rm cd} \frac{dT}{dr}\, ,
    \label{eq:cond_flux}
\end{align}
where $\lambda_{\rm cd}$ is the thermal conductivity of the material in question. 

Conduction via electron transport is important at higher pressures, though in the traditional energy transport model in giant planets, it is subdominant. Nevertheless, we include it for completeness and may find a role in some contexts. \citet{Hubbard1968} estimated the thermal conductivity {in the metallic H region} as
\begin{equation}
\lambda_{\rm cd} \sim 10^{8} \bigg(\frac{\rho}{1 \rm g\,cm^{-3}}\bigg)^{4/3} \,\rm ergs\,cm^{-1}\,s^{-1}\,K^{-1} \, ,
\end{equation}
{which is shown by the solid black line in the right panel of Figure \ref{fig:cond_fig}. \cite{Stevenson1977a} also obtained a similar expression in the metallic H region, and extended their calculations into the molecular H region, which is shown by the dashed green line.}
The dependence of $\lambda_{\rm cd}$ on the helium abundance is relatively small (but see \citet{preising2023}). \citet{French2012} carried out more refined calculations of material properties using ab initio molecular dynamics simulations, focusing on a hydrogen-helium-water mixture along Jupiter's supposed adiabat. They found that the thermal conductivity is predominantly electron-dominated in the metallic region, but decreases strongly at the transition zone to the molecular phase, where ionic contributions begin to dominate. This behavior is summarized {by the blue line} in the right panel of Figure \ref{fig:cond_fig} as a function of {density}. 

{Given that the \citet{French2012} calculations have {superceded} the prior estimates provided by \cite{Hubbard1968} and \cite{Stevenson1977a}, we now utilize the data from \cite{French2012}, 
% we use the functional form from \cite{Stevenson1977a} and modify it to match the \cite{French2012} data at a particular density,
and shift by 1.3 \citep{Zaghoo2017} in the current \texttt{APPLE} framework.} {We extrapolate linearly the thermal conductivity over density in regions outside the \citet{French2012} table, and when this extrapolation intersects the \citet{Stevenson1977a} formula in the molecular region at low densities we use it. Note that in the latter the thermal conductivity is a function of both density and temperature. We recognize that this approach is suboptimal, but comprehensive modern tables as a function of temperature and density throughout the thermodynamic range of relevance to giant planet evolution are not yet available. The recent sophisticated calculations by \citet{preising2023} for helium-rich conditions that might prevail in Saturn's interior are also available as an option for the thermal diffusivities employed in \texttt{APPLE}.} {We emphasize that \texttt{APPLE} can employ any formulation of the thermal conductivities and that the values shown in Figure \ref{fig:cond_fig} are merely the current defaults.}

%\footnote{There are no proper data on thermal conductivities for giant planet evolutionary purposes, which will require tables based on density, temperature, and composition.}

\subsection{Convection}
\label{sec:conv}

Convection is still at present, the dominant energy and species transport mechanism in planetary atmospheres and interiors. One of the most widely applied models for handling this complicated three-dimensional and turbulent process in a simplified spherical context is Mixing Length Theory (MLT) \citep{Bohm1958}. According to MLT, heat is transported by thermal parcels over characteristic length scales before dispersing into the surrounding fluid. The mixing length describes the average vertical distance traveled before a convective fluid element becomes thermally equilibrated with its surroundings. While the theory doesn't explicitly define this mixing length, it is commonly expressed in terms of the local pressure scale height ($ H_p = -\frac{dr}{d\ln P}$), and is defined as $l = \alpha H_p$, where $\alpha$ is a tunable constant (which we generally set equal to 1).

The convective flux in the MLT is conceptualized as the excess energy carried by a convective parcel, multiplied by its velocity. Mathematically, this is expressed as
\begin{equation}
\flux_{\rm conv} = \rho v_{\rm MLT} C_p \Delta T\, ,
\label{eq:convflux1}
\end{equation}
where $v_{\rm MLT}$ is its velocity, $C_p$ is the specific heat at constant pressure, and $\Delta T$ is the temperature difference between the fluid element and its environment. Under the assumption of small adiabaticity, the temperature variation can be expressed as
\begin{align}
    {\Delta T} &= \bigg[-\left(\frac{dT}{dr}\right) + \left(\frac{dT}{dr}\right)_{\rm ad}\bigg]\Delta r \\
    &= (\nabla - \nabla_{\rm ad})\frac{l}{2}\frac{T}{H_p}\, ,
\end{align}
where $\nabla = \frac{d \ln T}{d\ln P}$ and $\nabla_{\rm ad} = \frac{\partial\ln T}{\partial \ln P}\big|_S$. We assume the average distance traveled by a fluid element before mixing with its surroundings is $\Delta r = l/2$. The average velocity of the fluid element can then be calculated as \citep{Kippenhahn1990}:
\begin{equation}
v_{\rm MLT}^2 = g(\nabla - \nabla_{\rm ad})\frac{l^2}{8H_p}\, .
\label{eq:conv_vel}
\end{equation}
In the absence of composition gradients, we can express the temperature gradients in terms of the entropy ($S$) gradient as
\begin{align}
    \frac{dT}{dr} &= \frac{\partial T}{\partial P}\bigg|_S \frac{dP}{dr}  + \frac{\partial T}{\partial S}\bigg|_P \frac{dS}{dr}\\
     &= \frac{dT}{dr}\bigg|_{\rm ad}   + \frac{ T}{C_p}\frac{dS}{dr}
     \label{eq:TS}
\end{align}
Employing the definition of $H_p$ in equation \ref{eq:TS}, we get
\begin{equation}
\nabla - \nabla_{\rm ad} = -\frac{dS}{dr}\frac{H_p}{C_p}\, .
\label{eq:dsdr}
\end{equation}
The Schwarzschild criterion for convection \citep{Schwarzschild1958} is given by
\begin{equation}
    \nabla - \nabla_{\rm ad} > 0 
    \label{eq:schw_criterion1}
\end{equation}
and the corresponding convective flux\footnote{Note that, even though there is the $dY/dr$ term in the convective flux using the Schwarzschild criterion, the term in the bracket equals $\nabla - \nabla_{\rm ad}$. Also, the + indicates that only the positive value of the expression is used in the square bracket, otherwise, it is zero.} is \citep{Tejada2024}:
\begin{equation}
   \flux_{\rm conv} = \rho T \sqrt {\frac{ g l^4}{32C_p}} \left(\bigg[-\frac{dS}{dr} + \left(\frac{\partial S}{\partial Y}\right)_{P,T}\frac{dY}{dr}\bigg]_{+}\right)^{3/2}\, .
   \label{eq:sch_flux}
\end{equation}
However, if there is a composition gradient in the medium, the convective criterion is given by the Ledoux condition \citep{Ledoux1947} 
\begin{align}
    \nabla - \nabla_{\rm ad} - \frac{\chi_Y}{\chi_T}\frac{dY}{d \,{\rm log}\,P} &> 0 \\
    \implies \frac{dS}{dr}-\left(\frac{\partial S}{\partial Y}\right)_{P, \rho}\frac{dY}{dr}&<0\, ,
    \label{eq:full}
\end{align}
where $\chi_Y = \big(\frac{\partial \ln P}{\partial Y}\big)_{\rho, T}$, and $\chi_T=-\big(\frac{\partial \ln P}{\partial \ln T}\big)_{\rho,Y}$. The corresponding flux is approximated as
\begin{align}
\flux_{\rm conv} = \rho T \sqrt {\frac{ g l^4}{32C_p}} \left(\bigg[-\frac{dS}{dr}+\left(\frac{\partial S}{\partial Y}\right)_{P, \rho}\frac{dY}{dr}\bigg]_{+}\right)^{3/2}\, ,
\label{eq:ledoux_flux}
\end{align}
where we have generalized the flux here consistently to comport with the Brunt \citep{Kippenhahn1990} associated with the Ledoux condition, rather than the Brunt associated with the Schwarzschild condition. Numerically, we set $\flux_{\rm conv}=v_{\rm MLT}=0$ when a mass zone is stable to convection. 

Semiconvection has often been modeled \citep{Leconte2012,Mirouh2012} by introducing a parameter $R_{\rho}$, often referred to as the ``density ratio." In general, we may introduce a coefficient to the $\frac{dY}{dr}$ term in the convective flux:
\begin{equation}
    \frac{\partial S}{\partial Y} = (1 - R_\rho)\left(\frac{\partial S}{\partial Y}\right)_{P,T} + R_\rho\left(\frac{\partial S}{\partial Y}\right)_{P,\rho}\, ,
\end{equation}
such that the following criterion holds:
\begin{equation}
    R_\rho = 
    \begin{cases}
      0 & \text{Schwarzschild}, \\
      1 & \text{Ledoux},\\
     >0 \,\,\text{but} <1 & \text{``pseudo"-semiconvection}\, ,
    \end{cases}
\end{equation}
and we can mimic the thermal effect of semiconvection by varying $R_{\rho}$ \citep[see][]{Mankovich2016}. However, this effectively just shows how convection might vary between Ledoux and Schwarzschild and is not a good substitute for a more physical theory of semiconvection.
%where $R_c$ is a small number in the context of giant planet envelopes and is defined in the next section.

\subsection{Semiconvection}
\label{sub:semi}

Semiconvective heat transport is expected to occur in regions unstable to the Schwarzschild criterion but stable to the Ledoux criterion. Parameters said to govern energy transport in a semiconvective region \citep{Walin1964, Kato1966} are the Prandlt number ($\Pr$), the diffusivity ratio ($\tau$, also called the ``Lewis number''), and the density ratio ($R_0$), respectively, each defined as 

\begin{equation}
    \Pr \equiv \frac{\nu}{\kappa_T},
\end{equation} 

\begin{equation}
    \tau = \frac{\mathcal{D}}{\kappa_T},
\end{equation}

\begin{equation}\label{eq:R0}
    R_0 = \frac{\delta}{\phi}\bigg(\frac{\nabla - \nabla_a}{\nabla_Y}\bigg)\, ,
\end{equation}
where $\kappa_T$ and $\mathcal{D}$ are the thermal and compositional diffusivities, $\nu$ is the viscosity, $\nabla_Y$ is derived using eq. \ref{eq:full}, and

\begin{equation}
    \delta = \logpar{\rho}{T}{P}{Y},\ \phi = \logpar{\rho}{Y}{P}{T}\ ,
\end{equation}

\begin{equation}
    \nabla = \frac{d\log{T}}{d\log{{P}}},\ \rm{and}\ \nabla_a = \logpar{T}{P}{S}{Y}\, .
\end{equation}

In the context of giant planet envelopes dominated by hydrogen/helium mixtures, $\Pr$ and $\tau$ are approximately equal to $\sim$10$^{-2}$. Semiconvection occurs when $R_c < R_0 < 1$ \citep{Walin1964, Kato1966}, where

\begin{equation}\label{eq:Rc}
    R_c = \frac{\rm{Pr} + \tau}{\rm{Pr} + 1}\, .
\end{equation}

Regions where $1 > R_0 > R_c$ are unstable to semiconvection. Note that $R_0 = 0$ is the Schwarzschild condition and $R_0 = 1$ is the Ledoux condition. 

Several important studies of semiconvection in the regime relevant to giant planets have been conducted \citep{Rosenblum2011, Mirouh2012, Wood2013, Moll2016}. These authors explored the onset of semiconvection \citep{Rosenblum2011, Wood2013} which, though less efficient than full convection, nevertheless transports heat and mix matter through layered regions \citep{Mirouh2012, Moll2016}. 

\apple\ implements the empirical relations found by \citet{Mirouh2012} and \citet{Wood2013} to control the transport of heat and composition in semiconvective regions via the thermal and compositional Nusselt numbers, $\Nu_T$ and $\Nu_X$. In the case of non-layered convection, \citet{Mirouh2012} found that 

\begin{equation}
    \Nu_T - 1 \approx 0.75\bigg(\frac{\rm{Pr}}{\tau}\bigg)^{0.25\pm 0.15}\frac{1 - \tau}{\R - 1}\bigg(1 - \frac{\R - 1}{\Rc - 1} \bigg).
\end{equation}
In the case of layered convection, \cite{Wood2013} found that $\Nu_T$ and $\Nu_X$ can be approximated via empirical relations, 

\begin{equation}
    \Nu_T - 1 = A_T \rm{Ra}_T^a Pr^b
\end{equation}

\begin{equation}
    \Nu_X - 1 = A_X \tau^{-1} \rm{Ra}_T^c Pr^d\, ,
\end{equation}
where $A_T \approx 0.1$, $A_X \approx 0.03$, $a = 0.34 \pm 0.01$, $b = 0.34 \pm 0.03$, $c = 0.37 \pm 0.01$, and $d = 0.27 \pm 0.04$ \citep{Wood2013, Moll2016}. The parameter Ra$_T$ is the same modified Rayleigh number parameter used in stellar and planetary semiconvection studies \citep{Leconte2012, Spruit2013, ZaussingerSpruit2013,Wood2013, Nettelmann2015, Moore2016},

\begin{equation}
    \mathrm{Ra}_T = \frac{\alpha g l_H^4}{\kappa_T \nu}\bigg(\fullderiv{T}{r} - \fullderiv{T_{\rm{ad}}}{r}\bigg).
\end{equation}

In lieu of any exact knowledge of the various parameters on which the Nusselt numbers depend, \apple\ employs empirical relations with constant values of $\Pr=0.01$ and $\tau=0.01$. This simplifies the expression for the thermal Nusselt number found by \cite{Mirouh2012} to 

\begin{equation}
    \Nu_T \approx \frac{0.74}{\R - 1} + 0.85
\end{equation}
and if $c = 0.37$, $d = 0.27$, then the empirical relations of $\Nu_T$ and $\Nu_X$ found by \cite{Wood2013} become

\begin{equation}
    \Nu_T \approx 0.02\ \rm{Ra_T}(l_H)^{0.34} + 1 
\end{equation}

\begin{equation}
    \Nu_X \approx 0.86\ \rm{Ra_T}(l_H)^{0.37} + 1. 
\end{equation}
However, \texttt{APPLE} can vary such Nusselt values in semiconvective layers. The thermal Nusselt number is used to modify the heat flux due to conduction and is given by
\begin{align}
       \flux_s &= -\lambda_{\rm cd} (\Nu_T - 1)\bigg[\fullderiv{T}{r} - \fullderiv{T}{r}\bigg|_{\rm{ad}}\bigg] \\ 
       &= -(\Nu_T - 1) \frac{\lambda_{\rm cd} T}{C_P}\bigg[\fullderiv{S}{r} - \parderiv{S}{Y}{P}{T}\fullderiv{Y}{r}\bigg] 
\end{align}
so that the total flux in semiconvective regions becomes

\begin{equation}
    \flux_{\rm{tot}} = \flux_{\rm{cd}} + \flux_s \\ = -\lambda_{\rm cd} \bigg[\fullderiv{T}{r} + (\Nu_T -1)\bigg(\fullderiv{T}{r} - \fullderiv{T}{r}\bigg|_{\rm{ad}}\bigg) \bigg]
    \label{eq:semi}
\end{equation}
$$ = - \lambda_{\rm cd}\bigg[\fullderiv{T}{r} + (\Nu_T - 1) \frac{T}{C_P}\bigg[\fullderiv{S}{r} - \parderiv{S}{Y}{P}{T}\fullderiv{Y}{r}\bigg]\bigg].$$

\section{Mixing/Diffusion of Elements}
\label{sec:mixing}

The distribution of heavy elements plays a crucial role in understanding the formation and origin of the planets \citep{Helled2014, Helled2022}. Moreover, the mixing of chemical species by convection, microscopic diffusion, or phase separation (such as in helium rain) are important global processes to follow during a giant planet's evolution.  However, particularly in the case of phase separation and helium rain, the proper numerical approach to incorporate with some fidelity such multi-fluid/multi-phase physics into a spherical code is not yet clear. Presently, there are no cogent models of helium bubble growth, and in the case of non-convective mixing, a compelling model of droplet upwelling and settling \citep{Fortney2003, Pustow2016, Mankovich2016}. Therefore, in lieu of this, we have introduced an advective term into the general diffusion equation for a species such as helium that can phase-separate and rain. This establishes for raining species such as helium an asymptotic equilibrium scale, akin to a cloud deck, that captures the rain region in a manner consistent with the physical miscibility curves employed. In this spirit, we have crafted two schemes, both of which are fully consistent with whichever miscibility model is employed, and we describe them in \S\ref{subsec:helium_rain}. Next, we discuss the different types of mixing and their associated diffusion coefficients.

\subsection{Convective Mixing}

When convection is generally dominant in the interior such a region mixes matter very rapidly compared with evolutionary timescales (e.g.,  Kelvin-Helmholtz timescales). Given this, the chemical profile in convective layers can generally quickly flatten.  To accomplish this naturally for the evolution of any mass fraction $X_i=\{X,Y,Z\}$, we employ in such convective regions a diffusion equation:

\begin{align}
&\frac{dX_i}{dt} = \frac{\partial }{\partial M_r}\left(4\pi r^2 \rho \mathcal{D}\frac{\partial X_i}{\partial r}\right)\, ,
\label{eq:diffusion}
\end{align}
with the diffusion coefficient assumed to have the form
\begin{equation}
    \mathcal{D} = \Dmlt=\frac{1}{3}v_{\rm MLT} l\, ,
\end{equation}
where $v_{\rm MLT}$ is given by equation \ref{eq:conv_vel} and $l$ is the effective ``mean-free-path" for tubules, in this formulation set equal to the local mixing length. We note that this equation is non-linear since $\mathcal{D}\propto v_{\rm MLT}$, and is a complicated function of thermal profiles and thermodynamic quantities (\S\ref{sec:conv}).

\subsection{Semiconvective Mixing}
\label{sec:semi_mixing}
The transport of elements in a semiconvective region is governed by the compositional Nusslet number, $\Nu_X$. The Nusselt number will enhance a diffusion coefficient, $\mathcal{D}$ such that the semiconvective effective diffusion coefficient is $\mathcal{D}_{\rm sc} = \Nu_X\mathcal{D}$. The diffusion equation through a semiconvective layer is therefore

\begin{align}
&\frac{dX_i}{dt} = \frac{\partial }{\partial M_r}\left(4\pi r^2 \rho \mathcal{D}_{\rm sc}\frac{\partial X_i}{\partial r}\right)\, ,
\label{eq:diffusion2}
\end{align}
where $\Nu_X$ is either taken from the empirical relation of \cite{Wood2013} or left as a free parameter.

\subsection{Microscopic Diffusion}
\label{subsec:micro_diff}

Within quiescent regions, microscopic diffusion is generally slow and the associated diffusivities are small. In stars, gravitational settling proceeds at a glacial pace and has been studied quantitatively and in-depth for many decades. The mass fraction of helium ($Y$) in the Sun's atmosphere has taken its long lifetime to be depleted from $\sim$0.2725 to only $\sim$0.245 \citep{Peimbert2007}.  Nevertheless, this estimated primordial helium abundance is now compared with the \textit{Galileo} measurement of $Y_{\rm atm} \sim$0.234 in Jupiter's atmosphere \citep{vonZahn1998} to infer that there has been helium rainout even in Jupiter. To address this phenomenon, we have developed a set of schemes in a time-dependent and evolutionary context that, while not the final word, provides a useful and informative modeling context and methodology. When convection, semiconvection, and/or rain are not operative, we use a microscopic diffusivity, $\mathcal{D}_{\rm micro}$, that generally would result in a long diffusion timescale of order 1 Gyr across the whole planet, if dominant.  We now proceed to a discussion of our algorithms for handling helium rain with an advective-diffusive algorithm.

\subsection{Helium Rain}
\label{subsec:helium_rain}

Hydrogen-helium immiscibility was initially proposed to explain Saturn's luminosity excess by \cite{Stevenson1977a}. The settling of helium in the gravitational well would translate into the heating of the gas.  A gradual removal of helium from the planet's outer layer would occur in the immiscibility layer as helium droplets formed and rained down into the deeper interior. Recent theoretical miscibility diagrams for hydrogen/helium mixtures \citep{Lorenzen2011, Morales2009, Schottler2018} at Megabar (Mbar) pressures generally support this possibility, though these miscibility theories disagree in detail (\S\ref{subsec:H_He_misc}). To address this physics in an evolutionary code, we have developed two related, but different, approaches to handling it.

\subsubsection{Scheme A}
\label{subsec:schemeA}
In this framework, we model the helium rain as an advection/diffusion process and assume the same diffusion coefficient as pertains locally to any other species. Importantly, we use realistic hydrogen-helium phase diagrams (in $P-T$ space) and identify the region(s) at any given time where the planetary pressure-temperature profile intersects the miscibility curve. Within the planet, the intersection results in two pressure points, $P_1$ and $P_2$, where the helium phase separates and rains. $P_1$ represents the outermost pressure of immiscibility, whereas $P_2$ denotes the innermost pressure of immiscibility. The absence of $P_1$ results in no rainfall, whereas in the absence of $P_2$, we utilize the innermost pressure within the gaseous envelope. Therefore, the diffusion equation is modified to include an extra advective term for helium settling. In a convective region, we have:

\begin{align}
&\frac{dY}{dt} = \frac{\partial }{\partial M_r}\left\{4\pi r^2 \rho \left(\Dmlt \frac{\partial Y}{\partial r} + \bar{D}\frac{Y}{H_r} \right)\right\}\, ,
\label{eq:misc1}
\end{align}
where $H_r$ is an arbitrary scale parameter and
\begin{equation}
    \bar{D} =
\begin{cases}
\Dmlt & \text{if $P_1\leq P \leq P_2$} \\
0      & \text{otherwise}\, . \\
\end{cases}
\end{equation}
Notice that in the helium rain region, the $Y$ distribution would asymptotically settle to an exponential distribution with scale height $H_r$.

\subsubsection{Scheme B}
\label{subsec:schemeB}
In the second scheme, we add an advective term, but do not directly find a $P_1$ or $P_2$. At each pressure and temperature point, we identify the corresponding helium mass fraction $Y_{\rm misc}$, below which helium is immiscible and would rain out. To the diffusion equation (eq. \ref{eq:diffusion}), we add a modified advection term using $Y_{\rm misc}(P,T)$ at every point in the planetary envelope:
\begin{align}
&\frac{dY}{dt} = \frac{\partial }{\partial M_r}\left\{4\pi r^2 \rho \Dmlt \left(\frac{\partial Y}{\partial r} + \frac{{\rm max}(0,Y-Y_{\rm misc})}{\mathcal{H}_r}\right)\right\}\, .
\label{eq:misc2}
\end{align}
Here, $\mathcal{H}_r$ is not exactly the scale height as in scheme A, but sets the scale of the helium rain region. This characteristic scale is approximately set by the ratio of ${\mathcal{H}_r}$ to fractional $Y$ deficit. In addition, we ensure there is no rain below a pressure of one Mbar. The miscibility physics employed for both schemes is discussed in \S\ref{subsec:H_He_misc}. Moreover, we have the capability to shift miscibility curves to taste in the way some researchers have suggested may be necessary to comport with Jupiter and Saturn \citep{Fortney2003, Pustow2016, Mankovich2016, Mankovich2020}. 

\iffalse
\begin{figure*}
    \centering
    \includegraphics[scale=0.34]{figures/rotation_comp_newmisc.pdf}
    \caption{Comparison of the evolution of fiducial Jupiter models with and without rotation, for a core mass of  $M_c = 10 \, M_{\bigoplus}$ and a heavy element mass of $M_z = 5 \, M_{\bigoplus}$. We set $H_r=1\times10^8$ cm as default and use the rain model: scheme B to rain out helium. The top panel shows the helium fractions at different timestamps with the dotted line showing the Galileo constraint $Y=0.234$ \citep{vonZahn1998}. The middle and bottom panels show the corresponding entropy and temperature profiles, respectively. The gray solid line is the \cite{Lorenzen2011} miscibility curve $T_{\rm misc} = T(P, Y)$.}
    \label{fig:rotation_comp}
\end{figure*}

\begin{figure*}
    \centering
    \includegraphics[scale=0.34]{figures/eos_comp_newmisc.pdf}
    \caption{Comparison of the evolution of fiducial non-rotating Jupiter models with different equations of state, namely the SCvH \citep[][red]{scvh1995} and the CMS \citep[][blue]{Chabrier2019}, for the same set of parameters as used in figure \ref{fig:rotation_comp}. }
    \label{fig:eos_comp}
\end{figure*}

\begin{figure*}
    \centering
    \includegraphics[scale=0.34]{figures/totalZ_comp_newmisc.pdf}
    \caption{Comparison of the evolution of fiducial non-rotating Jupiter models with different heavy element mass fractions partitioned into the core and the envelope, for the same set of parameters as used in figure \ref{fig:rotation_comp}.}
    \label{fig:Z_partition}
\end{figure*}
\fi

\section{Microphysics}
\label{sec:microphysics}
\subsection{Hydrogen-Helium Equations of State}
\label{subsec:eos}

\begin{figure}
    \centering
    \includegraphics[scale=0.52]{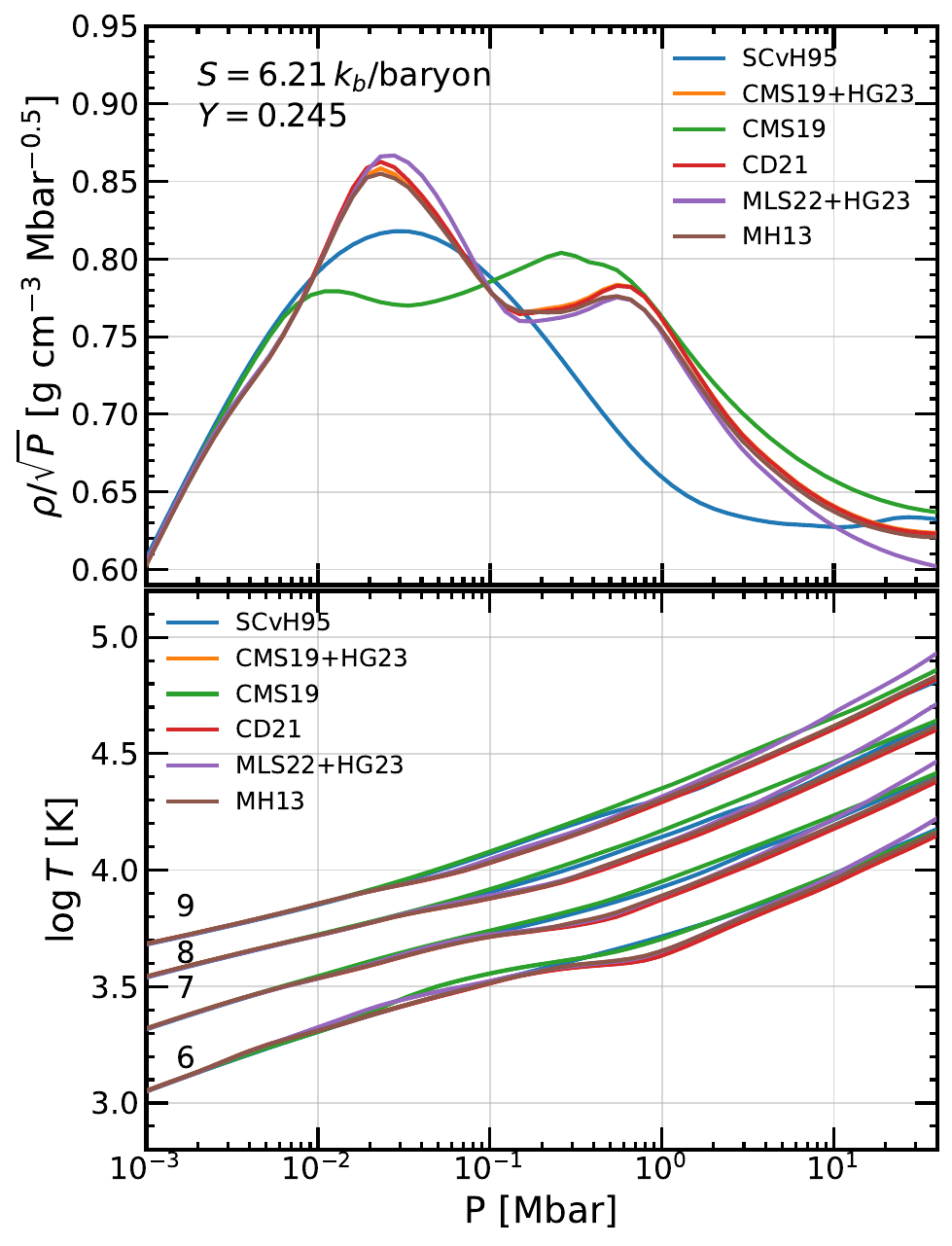}
    \caption{Reproduction of Figure 4 from \cite{Howard2023}, featuring corresponding labels for helium mass fraction $Y = 0.245$, and $S = 6.21$ kb/baryon. Bottom: Adiabatic profiles at various entropy values, measured in kb/baryon units. This behavior replicates findings illustrated in Figure 10 of \cite{Militzer2013} Under modest pressures of approximately 1 Mbar, non-ideal effects cause deviations from the equations of state proposed by SCvH95. Notably, these deviations occur in the same region where the demixing of hydrogen and helium is anticipated. (Figure inspired by that in \citet{Tejada2024}.)}
    \label{fig:EOS}
\end{figure}

The evolution of gas giant planets and brown dwarfs depends on the properties of hydrogen and helium (H-He) mixtures at high pressures \citep{Hubbard1970, Burrows1997, Fortney2003, Mankovich2020}. Early models of brown dwarfs and giant planets used the hydrogen-helium equation of state (EOS) computed by \cite{Scvh1995}. Under the ``chemical picture", this model estimated the gradual ionization of hydrogen and helium with increasing pressure and temperature and relied on the volume addition law to calculate the different EOS properties for an arbitrary helium mass fraction, $Y$. This chemical model has been used to evolve extra-solar giant planets, brown dwarfs, and low mass stars \citep{Burrows1997,Burrows2001} and to model giant planet evolution with helium rain \citep{Fortney2003,Mankovich2016}. However, as reviewed by \citet{Tejada2024}, this chemical EOS model could not incorporate partial ionization for species and non-ideal interaction corrections to the H-He entropy of mixing. 

Under the ``physical" picture, Density Functional Theory (DFT) with Molecular Dynamics (MD) was used to obtain these non-ideal interaction terms directly. These techniques produced a more physical H-He EOS, published by \citet[][MH13]{Militzer2013}. \citet{Hubbard2016}, \citet{Mankovich2020}, \citet{Militzer2024}, and \cite{Mankovich2021} then used this EOS to infer the structure and composition of Jupiter and Saturn at the present age. Nevertheless, this EOS was computed for only one helium mass fraction ($Y=0.245$) and requires extensions to accommodate other helium mass fractions. Physical EOSes published by \citet[][CMS19]{Chabrier2019}, \citet[][CD21]{Chabrier2021}, and \citet[][MLS22]{Mazevet2022} offer quantities that span the full range of thermodynamic space required for giant planet evolution, including a wide range of helium mass fractions, but require the use of the volume addition law \citep{Peebles1964, Chabrier1992} to accomplish the latter. The volume addition law is understood to be provisional and approximate and ab initio calculations for a wide range of helium fractions have yet to be performed and published. 

The EOS model of CD21 incorporates the non-ideal EOS of MH13 at $Y=0.245$. \citet[][HG23]{Howard2023} then calculated, using the MH13 and CD21 EOSes, these non-ideal effects for the CMS19 and the MLS22 EOSes for an arbitrary helium fraction. Currently, the CD21, CMS19, and MLS22 EOSes, with non-ideal terms calculated by HG23, are the most up-to-date models of H-He EOSes available for gas giant planet evolutionary calculations. Figure \ref{fig:EOS} compares these various equations of state along various adiabats and for a helium fraction of 0.245. These EOSes, coupled with updated H-He immiscibility curves, can be used to compute evolutionary models of Jupiter and Saturn with helium rain. \texttt{APPLE} allows one to employ not only the SCvH95 EOS, but also the new EOSes of MH13, CMS19, CD21, and MLS22. Importantly, \texttt{APPLE} can include the non-ideal entropy of mixing terms for the CMS19 and MLS22 EOSes via the EOS toolkit developed by \citet{Tejada2024}. 

\begin{figure}[ht]
    \centering
    \includegraphics[scale=0.56]{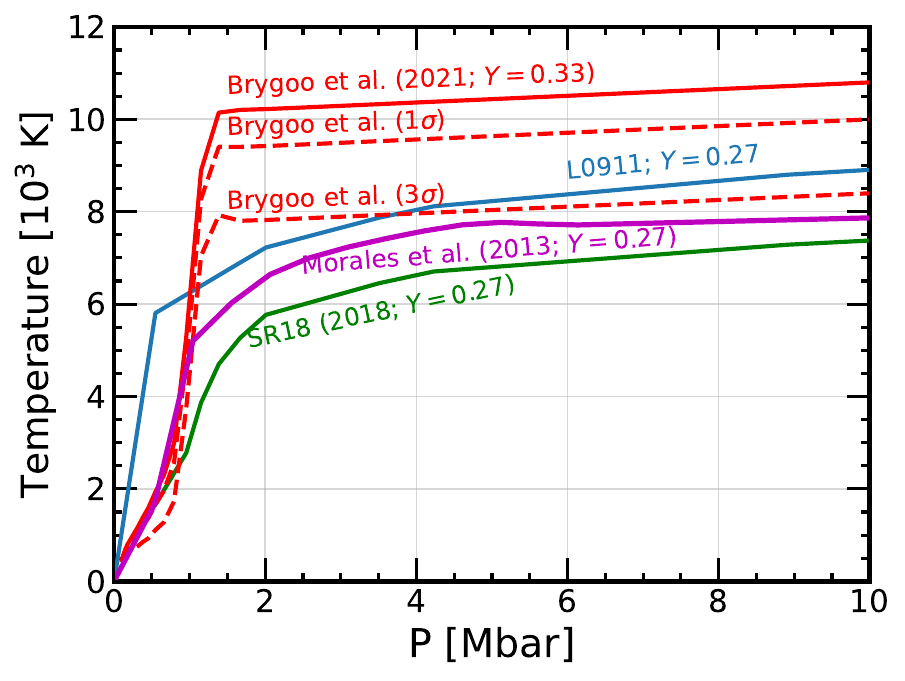}
    \caption{The various miscibility curves in T-P space are shown by different colors: \cite{Brygoo2021} (red), \citep{Lorenzen2009,Lorenzen2011} (blue), \cite{Morales2013} (purple), and \cite{Schottler2018} (green). The \cite{Brygoo2021} experiments found demixing temperatures far above the recent models of SR18 and L0911, indicating that our current understanding of these miscibility curves is only loosely constrained. Note that due to the sparseness of the \citet{Brygoo2021} data, the $\sigma$ error bar curves shown for them are notional. The curves are at fixed helium mass fraction ($Y = 0.27$). (Figure inspired by one found in \citet{Tejada2024}.)}
    \label{fig:misc}
\end{figure}

\begin{figure*}
\begin{minipage}{.33\linewidth}
\centering
\includegraphics[width=2.6in]{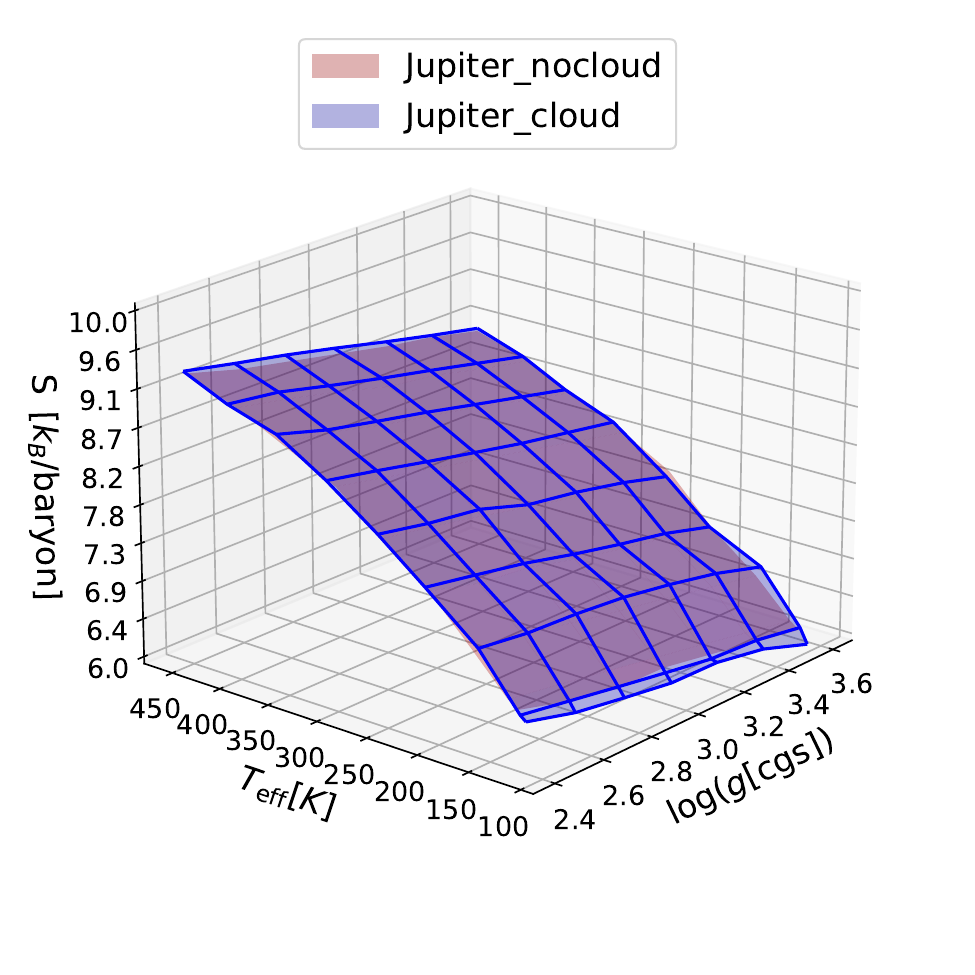}
\end{minipage}
%\hspace{0.01\linewidth}
\begin{minipage}{.33\linewidth}
\includegraphics[width=2.6in]{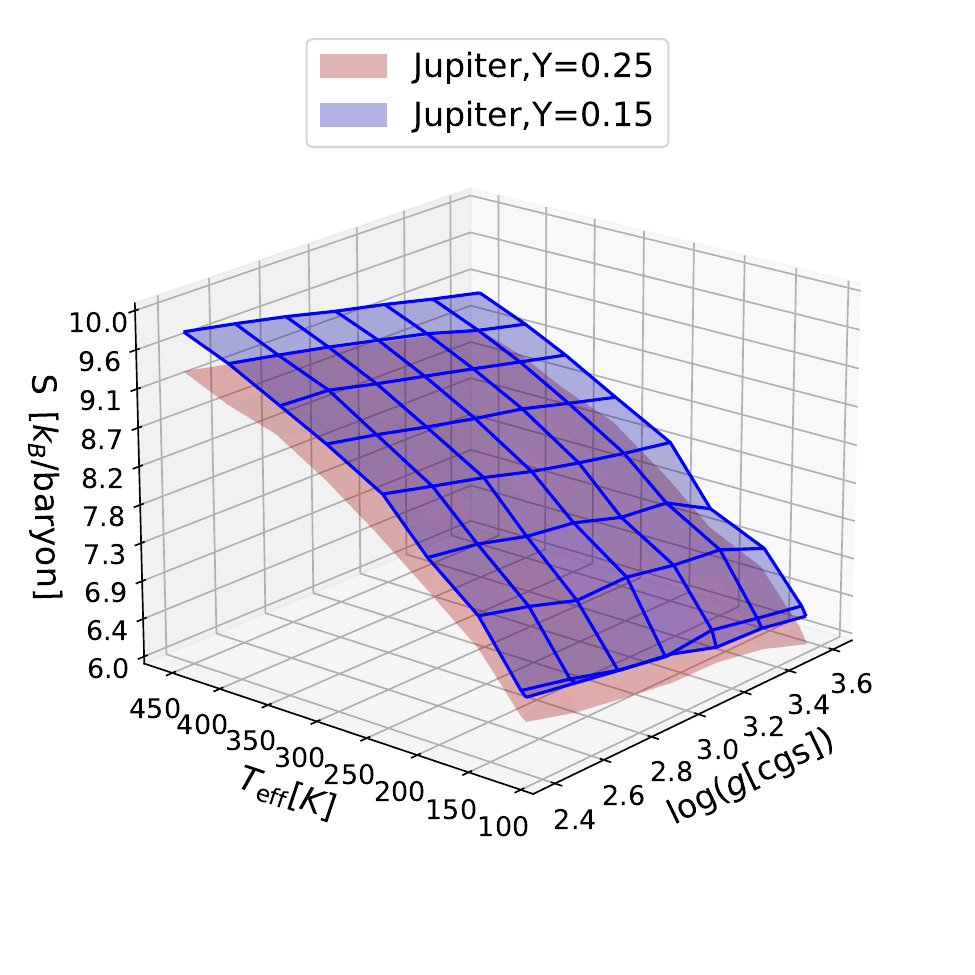}
\end{minipage}
\begin{minipage}{.33\linewidth}
\includegraphics[width=2.6in]{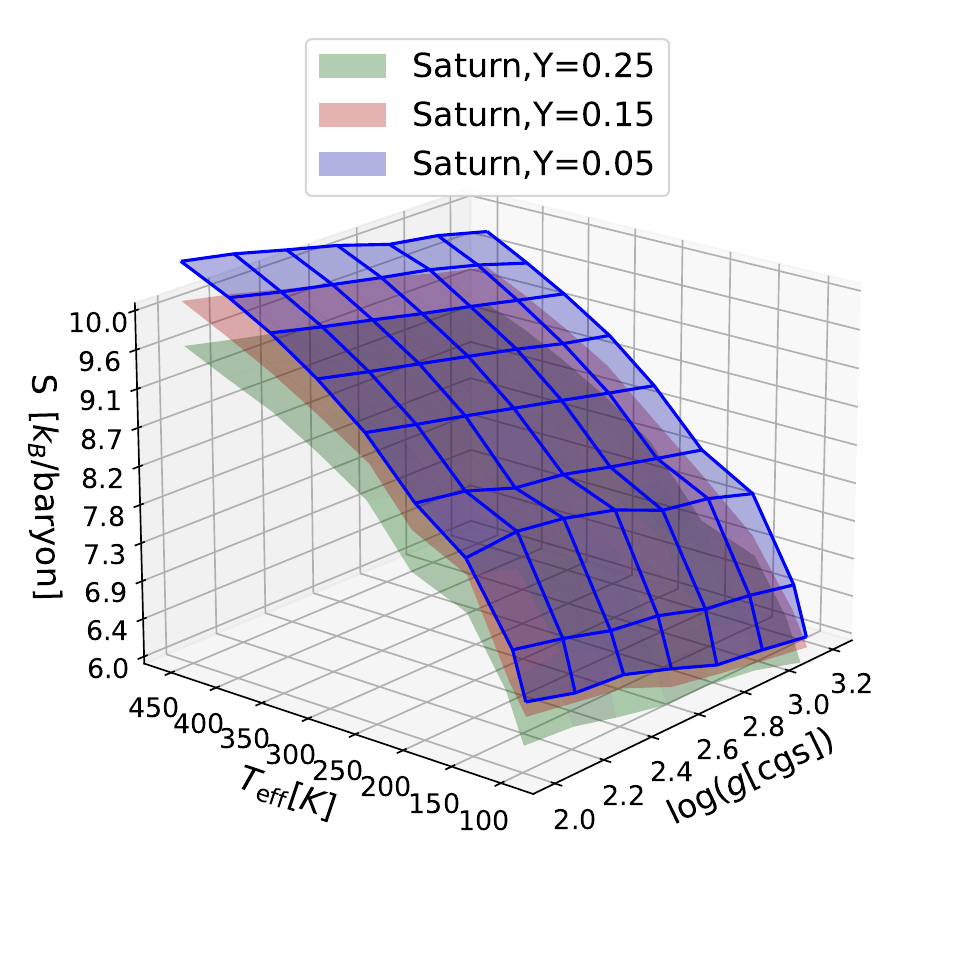}
\end{minipage}
    \caption{Effective temperature sheets for Jupiter and Saturn as a function of gravity and entropy. The left panel shows the effect of clouds in Jupiter. The middle and the right panels show the variation with helium fraction for Jupiter and Saturn, respectively. Jupiter models assume 3.16 solar metallicity and solar irradiation at a semi-major axis of 5.2 A.U., while Saturn models assume 5.0 solar metallicity and solar irradiation at a semi-major axis of 9.6 A.U.}
    \label{fig:atm_tables}
\end{figure*}
\subsection{Hydrogen-Helium immisciblity}
\label{subsec:H_He_misc}

To account for the excess flux observed in Saturn identified in early evolutionary calculations, \citet{Stevenson1975} and \citet{Pollack1977} proposed the existence of helium/hydrogen phase separation and helium rain. Early models of helium rain \citep{Fortney2003} with H-He phase separation curves calculated by \cite{Hubbard1985} found that helium rain could account for the hotter effective temperatures ($T_{\rm eff}$) observed in Saturn. The first phase separation models using MD techniques \citep{Pfaffenzeller1995} found that hydrogen-helium demixing temperatures were too cold for helium rain to exist in Jupiter or Saturn. Later DFT-MD-based demixing models for H-He mixtures by \citet[][L0911]{Lorenzen2009,Lorenzen2011}, \citet{Morales2013}, and \citet[][SR18]{Schottler2018} suggested hotter demixing temperatures at metallic hydrogen pressures. The models of L0911 and SR18 have since been used in gas giant planet evolutionary calculations. However, several researchers concluded that these models required shifts to the demixing temperatures for helium rain to in fact occur in Jupiter and Saturn \citep{Nettelmann2015, Pustow2016, Mankovich2016, Mankovich2020}. Figure \ref{fig:misc} compares examples of the various miscibility curves, along with an approximate Jupiter adiabat. As discussed in more detail in \cite{Tejada2024}, the H-He demixing temperatures remain highly uncertain at pressures relevant to gas giant planet interiors. This uncertainty is further exacerbated by the recent experiments of \cite{Brygoo2021}, who observed demixing temperatures far above the model demixing temperatures of L0911 and SR18. \texttt{APPLE} provides options to incorporate the L0911 and SR18 demixing curves as a function of pressure and helium abundance in ways that are fully consistent with their helium fraction dependencies (\S\ref{subsec:helium_rain}). It also provides a fit to the \citet{Brygoo2021} data and temperature-offset variants of the L0911 and SR18 curves.

\subsection{``Metal" Equations of State}

The gaseous envelopes and ``solid" cores of giant planets are thought to contain ``ices" and ``rocks", but the proportions and actual character of such constituents in either region are unknown.  \texttt{APPLE} provides a variety of mixtures that can be incorporated into models. For water, we use the EOS of \cite{Haldemann2020}, which incorporates different phases of water by collecting and connecting water EOSes from \cite{Feistel2006, Journaux2020, French2015, Wagner2002, Bollengier2019}, and \cite{Mazevet2019}. The pure water EOS can be combined with hydrogen and helium via the volume-mixing law, accounting for the entropy of mixing terms for each species via:

\begin{equation}
    S_{\rm{mix}} = S_{XY}(1 - Z)+ S_{H_2O}Z + S_{\rm{id}}\, ,
\end{equation}
where $S_{XY}$ is the hydrogen-helium mixture entropy, $S_{H_2O}$ the entropy of water, along with the water abundance (here denoted $Z$), and $S_{\rm{id}}$ is the ideal entropy of mixing. The ideal entropy of mixing of each component $i$ of $m$ systems is defined as 

\begin{equation}
    S_{\rm{id}} = k_b \bigg[N\ln{N} - \sum_{i=1}^m N_i\ln{N_i}\bigg]\, .
\end{equation}

This water EOS is used as our baseline for H-He-Z mixtures throughout the interior of our current models, but can easily be changed and generalized to accommodate any mixture. In cases where the H-He EOS contains non-ideal and ideal terms, like those found in the CMS19+HG23 and MLS22+HG23 EOSes, the ideal entropy of mixing of H-He is computed in advance and is then subtracted from the mixture to add the H-He-Z ideal entropy of mixing instead. This means that the H-He-Z mixture accounts for the three-component ideal entropy of mixing, but not the three-component non-ideal terms since those require ab initio mixture calculations. Currently, such mixture equations of state incorporating all interaction terms and non-ideal entropy terms are not available.

For post-perovskite and iron EOSes, \texttt{APPLE} incorporates those developed by Jisheng Zhang (private communication), which have been applied in rocky planet evolution models \citep{Zhang2022}. As a default for giant planet evolutionary calculations using \texttt{APPLE}, \cite{Tejada2024} have calculated various water/ice-perovskite-iron mixture tables, but these can easily be supplemented to incorporate any materials and mixtures. In summary, in \texttt{APPLE}, the metal content of the envelopes is assumed to be water, while the core is assumed to be made up of a mixture of water, silicates, and iron.  However, \texttt{APPLE} has been constructed to incorporate any mixture in either regime, as the need arises.  

%Future work on the EOS metal mixture will include other gases, such as Neon, whose depletion in Jupiter has been observed and considered evidence for helium rain in Jupiter at the present age \citep{Wilson2010}.

\section{Atmospheres}
\label{sec:atmospheres}
The effects of atmospheres on giant planet cooling were investigated by  \cite{Graboske1975}, while \cite{Hubbard1977} and \cite{Pollack1977} developed the first thermal evolution models of Jupiter and Saturn that coupled atmosphere boundary conditions and interior structures. Following this, \cite{Burrows1997} computed non-gray 1D atmosphere models using multi-frequency radiative transfer. These atmospheres were pre-calculated and tabulated. During an evolutionary simulation, the resulting table is interpolated to derive the internal flux temperature (${T_{\rm int}}$) and surface losses. This approach is more sophisticated than commonly employed in stellar and planetary evolution and is closely equivalent to performing detailed atmosphere boundary calculations at every evolutionary timestep 
\footnote{In the past \citep{Hubbard1977}, the quantity T10
(the atmospheric temperature at ten bars) was employed as a substitute for
planetary entropy and along with surface gravity was connected to the
effective temperature ($T_{\rm eff}$) through a simple power-law relation. This approach used rather simple atmospheric physics that didn't really capture the true opacities in the molecular atmospheres of giant planets and brown dwarfs. It was later superceded by the work of \citet{Burrows1997},
who calculated grids of atmospheres with a spectral code and interpolated
in these atmospheres during evolution.  The tables were in surface gravity
and the entropy at the base of the atmosphere. The base of
these atmospheres should be convective, and hence, isentropic, even if there
might be intermediate/convection-bounded radiative zones.
The thermal profiles and spectra of the latter were calculated with a detailed
``stellar" atmosphere code to depth that employed a full suite of molecular
opacities and put the atmosphere in chemical and radiative equilibrium. In
this way, one needn't calculate a real bounding atmosphere for every timestep
(an expensive proposition), but can interpolate in a grid of pre-calculated
atmospheres to obtain accurate $T_{\rm eff}$s (= ${T_{\rm int}}$ when the object is isolated) when performing cooling calculations. This algorithm is more sophisticated than that used in most of stellar evolution, but has since become the standard for subsequent giant planet and brown dwarf evolutionary calculations (see \citet{Burrows1997}, \citet{Burrows2001}; \citet{marley2021_sonora}; \citet{Baraffe1998}; \citet{Fortney2011}). For the \texttt{APPLE} code, we have gone a step further by including stellar irradiation of the planet in the detailed atmosphere calculation. Some researchers use the concept of T$_{eq}$ (equilibrium)
and set $T_{\rm eff}^4$ = ${T_{\rm int}^4}$ + T$_{eq}^4$.  For a variety of reasons, this ansatz is not particularly good, though a crude stopgap. As stated, however, in \texttt{APPLE} we use the Chen et al (2023). methodology, from which we can obtain $\mathbf{T_{\rm int}}$, $T_{\rm eff}$, the Bond albedo, and the emergent spectra in one consistent formalism.}. 
The tables map the entropy ($S$) at the base of the radiative layer and surface gravity $g$ to the internal brightness temperature ${T_{\rm int}}$ (eq. \ref{eqn:definition_Tint}) necessary to evolve brown dwarfs and giant planets, including Jupiter and Saturn \citep{Hubbard1999, Fortney2003}. In a similar fashion, \citet{Fortney2011} computed grids of radiative-convective model atmospheres for Jupiter, Saturn, Uranus, and Neptune over a range of ${T_{\rm int}}$ and $g$ that includes the influence of stellar/solar irradiation. These models were customized to the individual characteristics of each planet, such as appropriate chemical abundances. Their boundary conditions have been utilized in numerous planetary evolutionary simulations, such as \cite{Mankovich2016} and \cite{Mankovich2020}. However, the effect of clouds was not accounted for in these models.

\begin{figure*}
    \centering
    \includegraphics[scale=0.52]{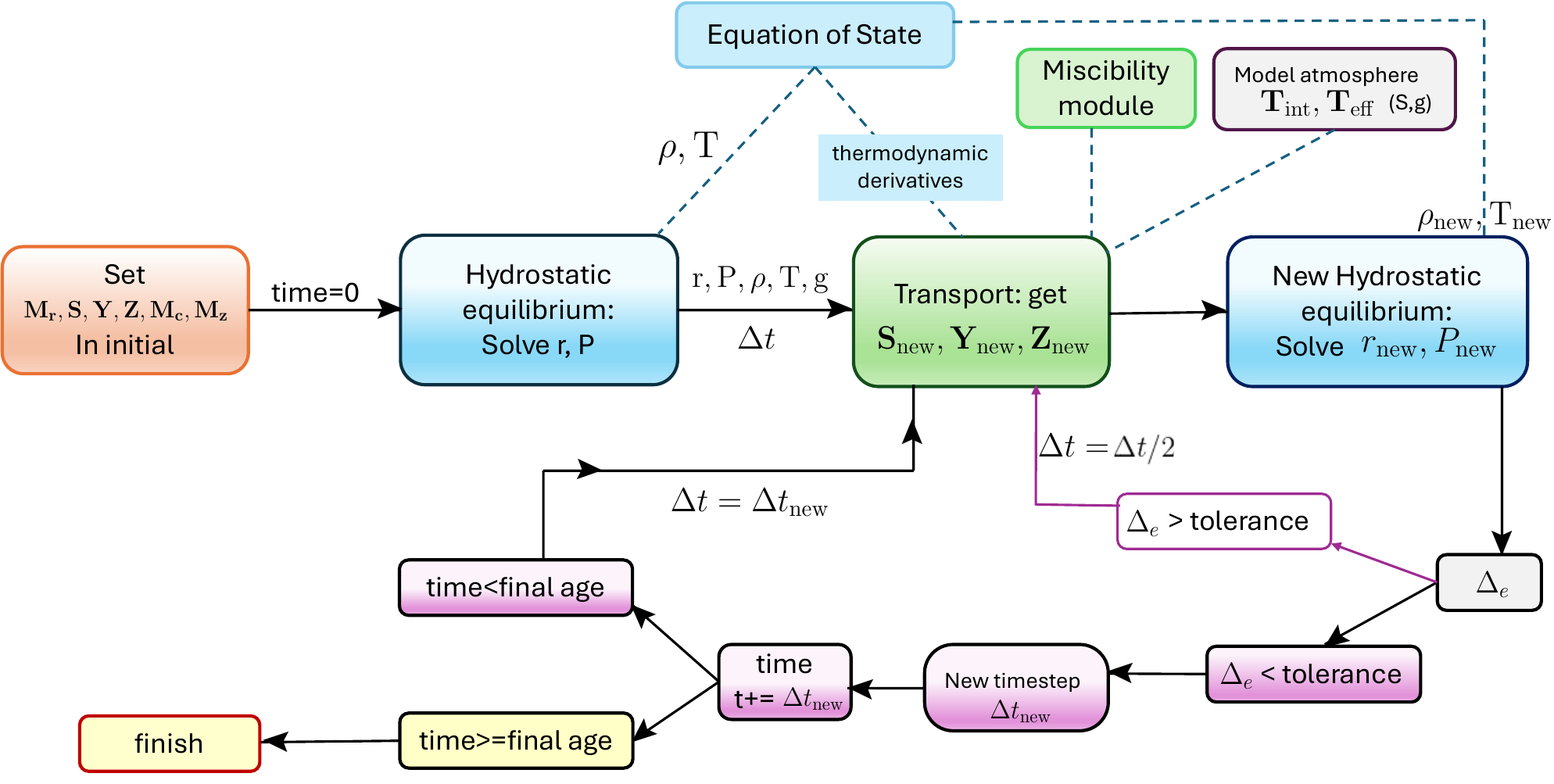}
    \caption{Flowchart of the \texttt{APPLE} code showing the different function calls. The $\Delta_e$ is calculated as the maximum of the fractional changes in density, temperature, entropy, and the helium fraction, while the tolerance is usually set to $1 \%$.}
    \label{fig:flowchart}
\end{figure*}

More recently, in the spirit of \citet{Burrows1997}, \citet{Chen2023} calculated a suite of boundary atmospheric models that include ammonia clouds with different particle sizes, irradiation, metallicity parameters, and the latest opacities \citep{Lacy2023}. For Jupiter evolutionary models, the fiducial cloud and irradiation parameters in \citet{Chen2023} have been calibrated to match the measured temperature structure and geometric albedo spectrum of Jupiter from \textit{Galileo} probe data \citep{vonZahn1992, Niemann1992, Young1996, Li2018} and can in principle be used for computing evolutionary models for giant exoplanets. \texttt{APPLE} incorporates the \citet{Chen2023} atmospheres as outer flux boundary conditions, which also allows it to generate both spectra and (for irradiated planets) reflection and Bond albedos as a function of time, but it can also incorporate the boundary conditions of \cite{Fortney2011} or those of \cite{Burrows1997} at the choice of the user. \citet{Chen2023} included various improvements to past implementations, such as updated chemistry \citep[e.g.,][]{Loders2002, Lacy2023}, non-gray opacities \citep{Sharp2007, Freedman2008}, cloud formation \citep{Burrows2006b}, and the effects of stellar irradiation \citep{Sudarsky2000,Baraffe2003,Burrows2006a,Fortney2011}. The boundary condition table maps both $S$ and $g$ to both $T_{\rm eff}$ and $T_{\rm int}$ (which deviates from $T_{\rm eff}$ when heating from irradiation is significant). Figure \ref{fig:atm_tables} depicts sample tables. The \texttt{atm\_bc} module in \texttt{APPLE} allows one to switch between \cite{Burrows1997}, \cite{Fortney2011}, and \cite{Chen2023} boundary conditions. \texttt{APPLE} offers the option to interpolate between purpose-built tables at a variety of atmospheric helium abundances ($Y_{\rm atm}$). This is important in the context of significant helium depletion of the atmosphere due to helium rain. The general methodology described here for handling atmospheres is the default in \texttt{APPLE} for simulating the evolution of giant exoplanets.

\section{Calculational Algorithm}
\label{sec:methods}
In addressing the structure and evolution equations \ref{eq:1}-\ref{eq:4}, we adopt the Henyey formalism \citep{Henyey1964}, which employs the Lagrangian coordinate $M_r$ in spherical symmetry, rather than the radius $r$. Consequently, all variables, including radius and density, are expressed as functions of mass coordinates and time, denoted as $r(M_r,t)$ and $\rho(M_r,t)$, respectively. Our calculations involve using an operator split technique, where we first compute the hydrostatic equilibrium for our initial conditions. Then, we advance the time and fully implicitly update the entropy, using the energy equation, and the element mass fractions, using the mixing/diffusion prescriptions (\S\ref{sec:mixing}), after which we loop back to put the structure into hydrostatic equilibrium at the new entropies and mass fractions.

The initialization process involves setting user-defined parameters such as the total planetary mass $M$, core mass $M_c$ in Earth masses, envelope heavy element mass $M_z$ in Earth masses, initial $Y_0$ and $Z_0$ profiles, and an initial entropy ($S_0$) profile. We can incorporate diverse materials within the core and can employ various H-He EOSes in the envelope \citep{Scvh1995, Chabrier2019, Chabrier2021}. These are integrated into the hydrostatic module, which uses an iterative two-dimensional Newton-Raphson scheme for computations, i.e.
\begin{align}
        \vec{F}(\vec{x}+\delta \vec{x}) = \vec{F}(\vec{x}) + \frac{d\vec{F}}{d\vec{x}}\delta \vec{x} = 0
        \label{eq:newton1}\\
        \vec{x}_{i+1} = \vec{x}_i - [J(\vec{x}_i)]^{-1}\vec{F}(\vec{x}_i) 
        \label{eq:newton2}\\
        \vec{x}_{i+1} - \vec{x}_i \leq \epsilon \, ,
        \label{eq:newton3}
\end{align}
where $\vec{F}$ takes in equations \ref{eq:1} and \ref{eq:2} for the hydrostatic structure, $\vec{x} = \{r,P\}$. $\vec{J}$ is the Jacobian matrix, with components $J_{ij} = \partial{F_j}/\partial{x_i}$, and  $\epsilon$ is the Newton-Raphson tolerance. The hydrostatic solver requires an initial guess for the pressure ($P_g$) and the radius ($r_g$) profiles which are obtained for the initial model at time zero using a $4^{th}$-order Runge-Kutta integration scheme. Newton's method has the advantage of \textit{quadratic convergence}, which means it converges quickly when the initial guess is close to the solution. At each iteration, $i$, the density and temperature profiles are calculated from the EOS module, which has been structured to use $S, P, Y, {\rm and}\ Z$ as independent variables.  

Next, a miscibility scheme needs to be chosen. Default options are the L0911 \citep{Lorenzen2009, Lorenzen2011}, SR18 \citep{Schottler2018}, and  Brygoo21 \citep{Brygoo2021} miscibility curves, but any other miscibility theories can easily be incorporated, as can shifts in temperature to the default miscibility models. Under helium rain scheme A, the miscibility module returns two pressure points $P_1$ and $P_2$ where the $P-T$ profile of the planet intersects the miscibility curve. If no intersection is found, the module returns \texttt{None}. Conversely, scheme B returns $Y_{\rm misc} (P, T)$. 

A small initial timestep $\Delta t$ is selected before the start of the evolution. With the planet in hydrostatic equilibrium, entropy $S$ and the element fraction $Y$ are updated at time $\Delta t$. The helium rain algorithm is chosen using the miscibility scheme. Temperature gradients in the radiative and conductive fluxes are treated implicitly where the temperature is expanded, via
\begin{equation}
  T_{\rm new} = T_{\rm old} + \left( \frac{\partial T}{\partial S}\right)_{\rho, Y}\Delta S + \left(\frac{\partial T}{\partial Y}\right)_{\rho, S}\Delta Y \, ,
\end{equation}
(dropping for the moment the $\Delta Z$ terms). The convective flux is expressed as an entropy gradient such that the evolution tends to make the entropy flat (in fact, slightly ``superadiabatic") in regions with uniform compositions. To calculate the boundary energy flux ($\sigma T_{\rm int}^4$), $T_{\rm int}$ is interpolated from the atmosphere grid using the surface gravity ($g$) and entropy. The thermal evolution equation \ref{eq:3b}, along with the species diffusion equation \ref{eq:4}, are discretized into algebraic equations and solved fully implicitly in time using, again, the iterative Newton-Raphson method (equations \ref{eq:newton1}-\ref{eq:newton3}). The tolerance ($\epsilon$) for the Newton-Raphson convergence is set at $10^{-6}$. Almost all thermodynamic derivatives are treated explicitly, and evaluated using the planet profile at the current time. The exception is the derivative entering into $\mathcal{D}$ via $v_{\rm MLT}$ (eq. \ref{eq:conv_vel}), which is updated every Newton-Raphson iteration. This improves numerical stability. To tackle this system of nonlinear algebraic equations, and the hydrostatic equations, we utilize a sparse matrix solver from the linear algebra package in \textit{SciPy}.

\begin{algorithm}[H]
 \SetKwInOut{Input}{Input}
 \SetKwInOut{Output}{Output}
 \Input{Set initial parameters: $M, M_c, M_z, \rm{tolerance}, \epsilon$\\and initial values for $r_{\rm g}, P_{\rm g}, S_o, Y_o, Z_o, t=0$ \par}
 \vspace{0.1cm}
 $r_o, P_o, \rho_o, T_o, g$  = hydrostatic ($r_g, P_g, S_o, Y_o, Z_o, \epsilon$) \par
 \vspace{0.1cm}
 Set $\Delta t$ (taken to be small) \;\par
 \SetAlgoLined
 \vspace{0.1cm}
 \While{t $\leq$ {\rm final\_age}}{
 Call miscibility module\;\par
 \vspace{0.1cm}
 Get $T_{\rm int}(S_o,g_o)$ from atmosphere boundary condition\par
 \vspace{0.1cm}
  $S_{\rm n}, Y_{\rm n}, Z_{\rm n}$ = transport($r_o, P_o, \rho_o, T_o, S_o, Y_o, Z_o, T_{\rm int}, \Delta t, \epsilon$)\;\par
  \vspace{0.1cm}
 $r_{\rm n}, P_{\rm n}, \rho_{\rm n}, T_{\rm n}, g =$ hydrostatic ($r_o, P_o, S_n, Y_{n}, Z_{n}, \epsilon$)\;\par
 \vspace{0.1cm}
  Calculate $\Delta_e$\;\par
  \vspace{0.1cm}
  \eIf{ $\Delta_e > {\rm tolerance}$}{
   Repeat steps 6 and 7 with $\Delta t= \Delta t/2$
  }{
   Calculate new timestep: $\Delta t = \Delta t_{\rm new}$\par
   Advance time: $t+=\Delta t$\par
   Store all variables \par
   Set: \\$S_o=S_n\\ T_o=T_n\\ r_o=r_n\\ P_o=P_n\\ Y_o=Y_n\\ Z_o=Z_n$\par
   \textbf{continue}\par
   }
}
\Output{Save final values}
\caption{Main Algorithm for \texttt{APPLE}}
\label{algorithm}
\end{algorithm}

With the new entropy $S_{n}$, and new species fractions $Y_n$, $Z_n$, a new hydrostatic equilibrium is calculated. Following this, we estimate the maximum fractional changes of $S, Y, Z, \rho, T$,
\begin{equation}
\Delta_e = {\rm max}\left(\frac{|\Delta S|}{S},\frac{|\Delta Y|}{Y}, \frac{|\Delta Z|}{Z}, \frac{|\Delta \rho|}{\rho}, \frac{|\Delta T|}{T}\right) \, ,
\label{eq:time_tolerance}
\end{equation}
and compare with the user-defined timestep tolerance. If the change is below the time tolerance, we allow the timestep to grow according to
\begin{equation}
    \Delta t_{\rm new} = {\rm min}(\Delta t_{\rm old}*{\rm min}({\rm tolerance}/{\Delta_e},2),{\rm max\_step}) \, ,
\end{equation}
where max\_step is chosen to be 20 Myr. If $\Delta_e$ exceeds the time tolerance, the timestep is halved and subsequent steps (i.e., the thermal and hydrostatic updates), are recalculated until $\Delta_e \leq \rm tolerance$. Lastly, time is advanced and compared with the planet's final age. A complete flowchart of the code is given in Figure \ref{fig:flowchart} and the outline is summarized in Algorithm \ref{algorithm}.

\begin{figure*}
    \centering
    \includegraphics[scale=0.44]{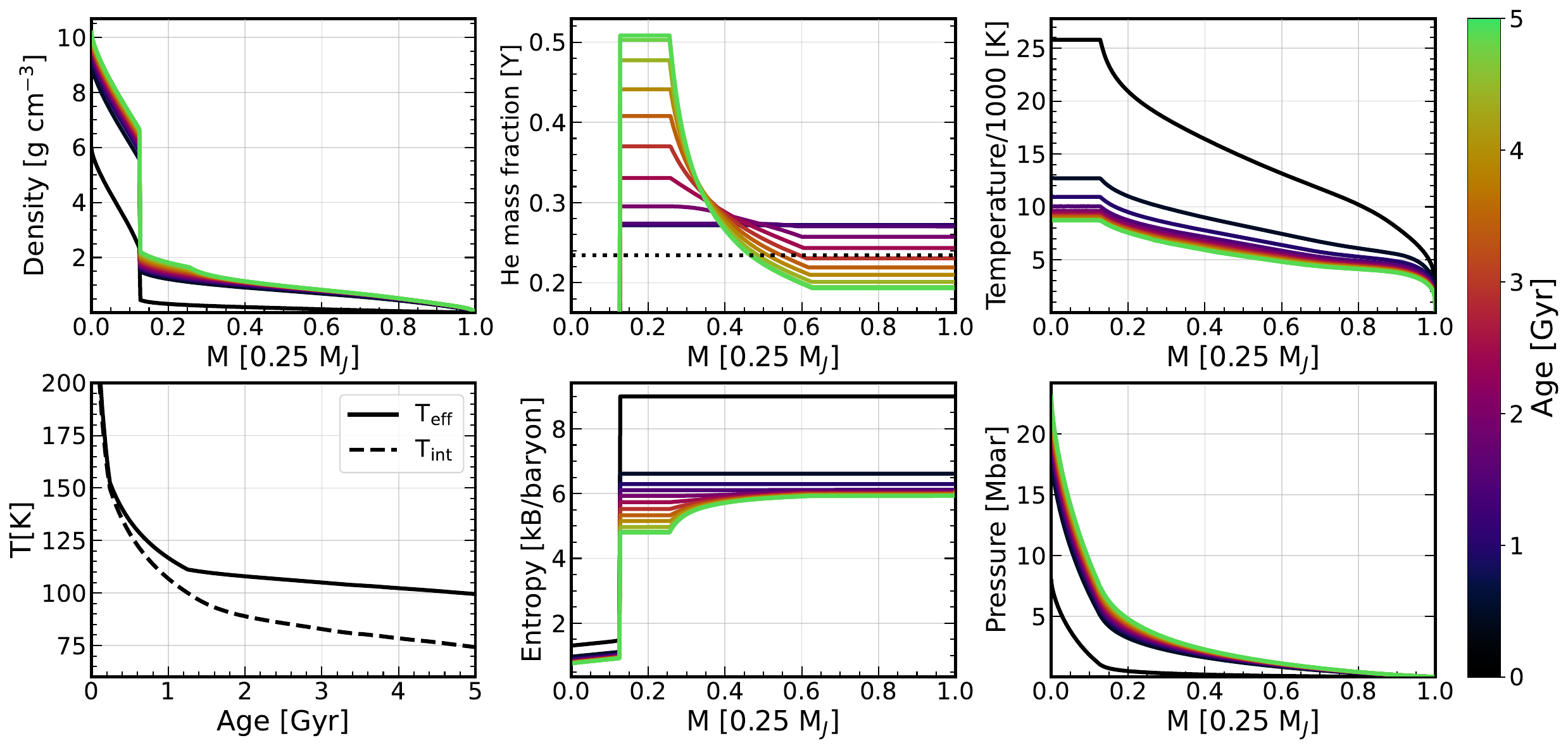}
    \caption{Evolutionary profiles for a non-rotating 0.25 $M_J$ planet characterized by a core mass of $10 M_{\oplus}$, and a H-He envelope uniformly mixed with heavy elements, with a mass of $5 M_{\oplus}$. The CMS19+HG23 EOS is utilized for the H-He mixture, while the L0911 miscibility curve with no temperature shift is employed to model helium rain. Various parameters including density, helium mass fraction, temperature, pressure, and entropy are displayed in panels moving clockwise from the top-left to bottom-middle. The planet's effective temperature ($T_{\rm eff}$) and internal temperature ($T_{\rm int}$) are presented in the bottom-left panel. The dotted line in the top-middle panel shows the \textit{Galileo} constraint $Y_{\rm atm}\sim0.234$ \citep{vonZahn1998}. We have used the 5 solar metallicity Saturn atmospheric boundary condition. The color scale indicates the different ages in Gyr. Heat transport via convection uses the Schwarzschild criterion. The helium fraction in the core is zero.}
    \label{fig:025mjup}
\end{figure*}

\subsection{Energy Conservation}
\label{sec:energy_cons}
Conventionally, the total energy is the sum of the internal and gravitational potential energies. This means at any time $t$, we have
\begin{equation}
    E_{\rm tot}(t) = \int_0^{M_p} U(M_r,t) dM_r - \int_0^{M_p} \frac{G M_r}{r(M_r,t)} dM_r\, ,
    \label{eq:energy_conservation1}
\end{equation}
where here we have dropped for convenience the rotational kinetic energy term (associated with the last term in eq. \ref{eq:1}).
Conservation of energy means that the change in $E_{\rm tot}$ over time $t$ must be compensated by the total radiation losses from the surface:
\begin{equation}
    E_{\rm tot}(t) - E_{\rm tot}(0) + \int_0^t 4\pi R_p^2 \sigma T_{\rm int}^4 dt = 0.
    \label{eq:energy_conservation2}
\end{equation}
 In our code, although we implement the \texttt{tds} form of the energy equation (eq. \ref{eq:3b}) during transport computations,  we utilize equation (\ref{eq:energy_conservation1}) for assessing energy conservation. Across most simulations, with a time tolerance of 1\% and 500 spatial zones, we observe energy conservation of approximately 0.8\% compared to the total radiation energy lost within 5$-$10 Gyrs. This value improves with finer spatial resolution and stricter time tolerance settings, but we find this level of conservation adequate for our detailed simulations. We note that this quantity is rarely tracked in other extant codes, but is a useful index of code fidelity.

\section{Example simulations}
\label{sec:simulations}
\begin{figure*}
    \centering
    \includegraphics[scale=0.445]{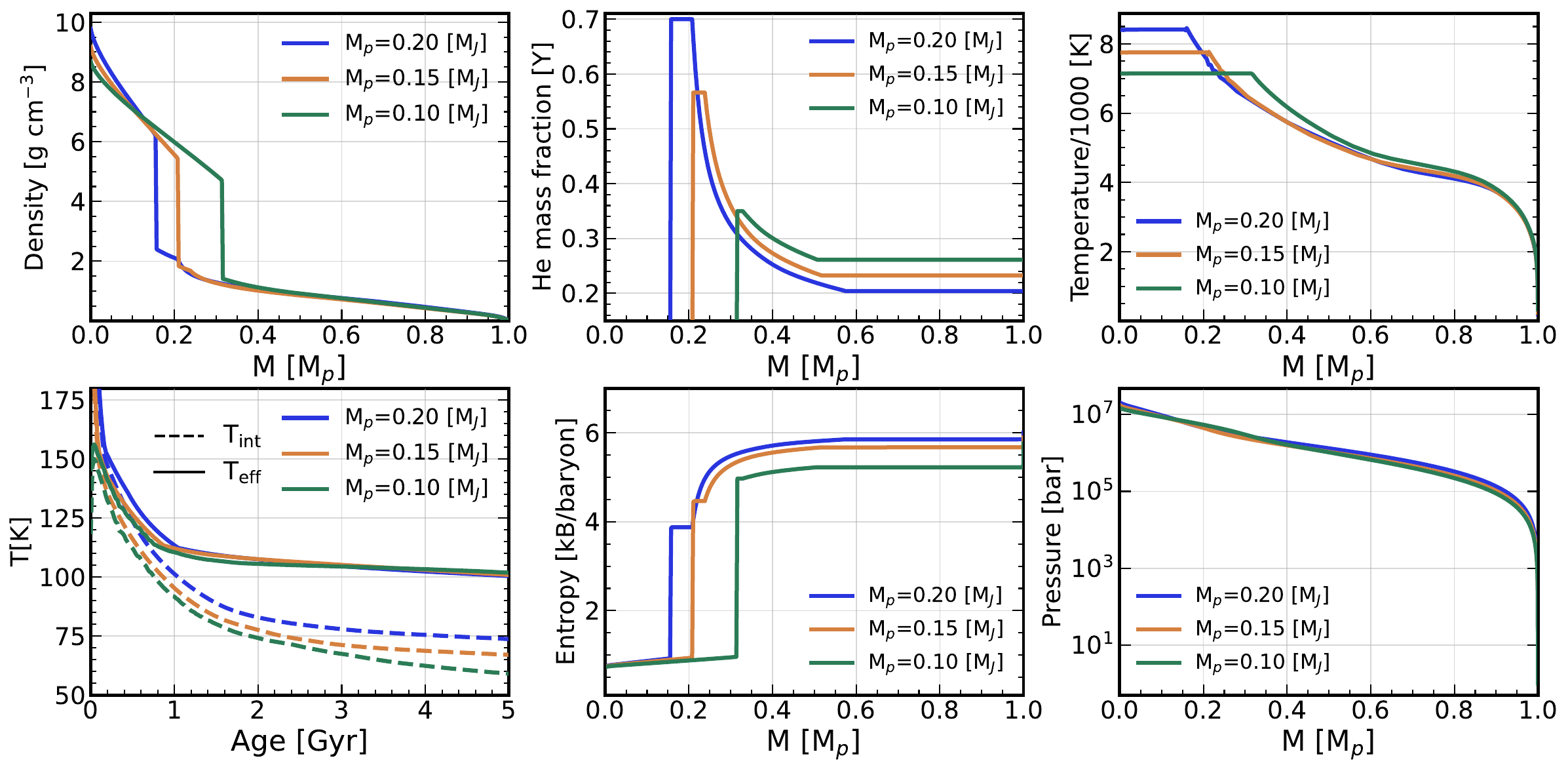}
    \caption{A comparison between different thermodynamic state variables: helium fraction, effective temperature ($T_{\rm eff}$), and internal temperature ($T_{\rm int}$) at the 5 Gyr mark for three distinct planet masses of $M_p = 0.2 M_J$ (blue), $M_p = 0.15 M_J$ (orange), and $M_p=0.1 M_J$ (green). All models have $M_c=10 M_{\oplus}$, and $M_z=5 M_{\oplus}$. The model with $M_p=0.2 M_{J}$ undergoes more helium depletion due to a larger miscibility region. The $T_{\rm eff}$ is, however, indistinguishable and is sensitive to the atmospheric boundary condition employed.}
    \label{fig:comparisons}
\end{figure*}
In this section, we provide example simulations for a variety of model setups and giant planets of different masses. Rotation is not included and each model incorporates a dense core, composed of $50\%$ water ice, $33\%$ post-perovskite, and 17\% iron mixture with a total mass $M_c=10$ Earth masses (M$_{\oplus}$). Surrounding this core is an H-He envelope, enriched with water ice, accounting for a heavy element mass ($M_Z$) of 5 M$_{\oplus}$ uniformly distributed in the envelope. For the water ice, we employ the AQUA EOS, and for the H-He mixture, the CMS19+HG23 is used. Note that this EOS has non-ideal entropy correction terms as incorporated by \cite{Howard2023}. The initial helium mass fraction in the envelope is set at $Y=0.272$ \citep{Lodders2019}. It is not well understood what initial compositional and entropy profiles arise due to planet formation \citep{Helled2014, Lozovsky2017, Helled2017}. Therefore, for these test runs we set our initial entropy to $S_0=9$ k$_B$/baryon. This conforms to a hot-start initial condition \citep{Marley2007,spiegel2012}, where the planet is puffed-up and has a radius $\sim{2}$ times the radius of Jupiter ($R_J$). For the low-mass planet model tests, we use the Saturn atmospheric boundary condition to calculate the outward flux. That table is constructed for $5\times$ solar metallicity and is available for three helium fractions, i.e. 0.25, 0.15, and 0.05, in which we interpolate to derive $T_{\rm int}$ for an instantaneous outer helium fraction $Y$. The heavy element mass fraction, $Z$, is held constant in time. In all of these example models, the evolution uses our helium rain scheme B outlined in \ref{subsec:schemeB}, using the L0911 miscibility curve with no temperature shifts. The value of the parameter $\mathcal{H}_r$ is set equal to $=10^8$ cm. We use 500 mass zones by default. By implementing a time step tolerance of 1\%, we achieve energy conservation close to 0.8\% and note that minor adjustments to this tolerance do not markedly alter the outcomes. The data are saved every 20 Myr during the evolution. We emphasize that the models discussed in this section are not our preferred models for the objects studied and that we plan soon to publish these in upcoming papers.

\subsection{Evolution of Low-Mass Giant Planets}
We describe here the evolution of a model $0.25$ M$_J$ giant planet and assume the Schwarzschild condition for convective stability within the planet. Figure \ref{fig:025mjup} shows the evolution with time of several key metrics relative to the planet's normalized mass, including density, helium mass fraction, temperature, entropy, pressure, and internal and effective temperatures. The color scale represents the planet's age in Gyrs. For these models, we embedded core masses containing roughly 12.5\% of the total planetary mass. The cooling process of the planet is evident from its temperature profiles, indicating a decrease in temperature and entropy over time. Heat transfer from the core is modeled purely via conduction, assuming a conductivity $\lambda = 10^{12}$ ergs cm$^{-1}$ s$^{-1}$ K$^{-1}$. In this way, the thermal diffusion timescale in the core remains small, ensuring it remains nearly isothermal at every time step. This can be seen in the top-right panel, where the temperature profiles in the core remain flat. Rather than assuming constant temperatures in the planet's core \citep{Baraffe2008, Lopez2014}, we allow in this model the core and mantle to couple thermally. The pressure profiles are continuous (bottom-right panel) with central pressure $P_c$ reaching almost 24 Mbar at the end of our simulation. 

\begin{figure*}
    \centering
    \includegraphics[scale=0.43]{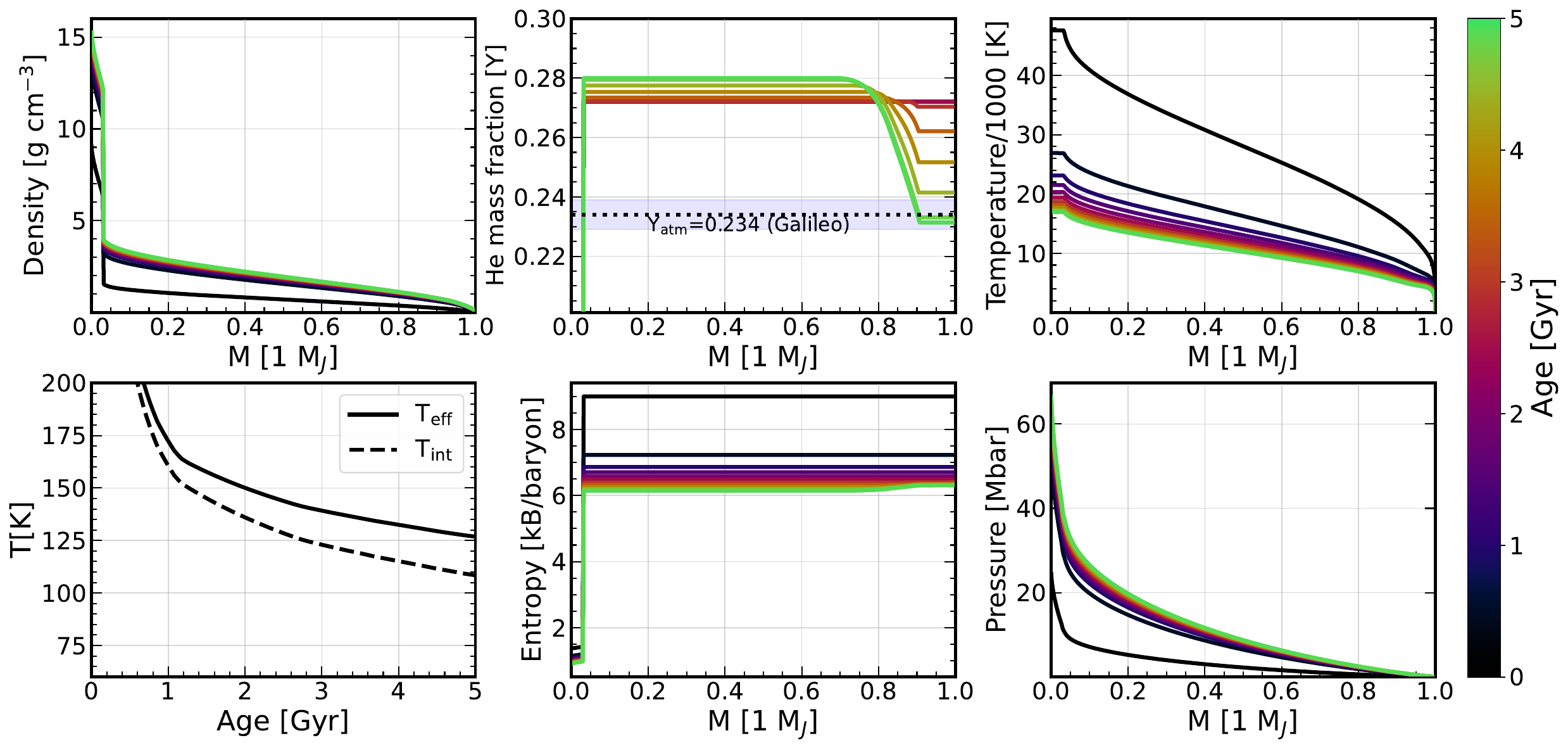}
    \caption{Same as Figure \ref{fig:025mjup}, but evolutionary profiles for a 1 $M_J$ model. We use the \cite{Chen2023} atmospheric boundary condition to calculate the outward-going flux. The dotted line in the top-middle panel shows the \textit{Galileo} constraint with the shaded region showing the uncertainty in the measurement \citep{vonZahn1998}.}
    \label{fig:1mjup}
\end{figure*}
As indicated in Figure \ref{fig:025mjup}, the planet's early compositional profile is homogeneous until its temperature-pressure profile intersects the L0911 miscibility curve at around 1.2 Gyr. At this point, helium starts differentiating from hydrogen, rains out, and redissolves in metallic regions at higher pressures. As a result, the $Y$ profiles increase toward the core, and the fraction in the outer envelope drops (top-middle panel). We observe flat $Y$ profiles below and above the demixing region due to the convective mixing of material on short time scales there, around 100 years, compared with our simulation time step of 20 million years. As previously mentioned, there is no helium diffusion into the core. This model experiences significant helium rain and its immiscibility region grows with time, as seen in previous studies \citep{Fortney2003, Pustow2016, Mankovich2016}. The atmospheric helium abundance reaches $Y_{\rm atm}\sim 0.19$, while the helium fraction below the immiscibility region becomes $\sim$0.5, conserving the total mass of helium. As illustrated in the bottom-middle panel, this phenomenon generates heat and results in a slowing of the decrease in the entropy of the planet's outer envelope.  As a result, the greater the depletion of helium, the greater the heating effect and the greater the effect on the planet's cooling. Notably, as depicted in the bottom-left panel, the evolution of the effective temperature ($T_{\rm eff}$) and the internal temperature ($T_{\rm int}$) over time presents an intriguing aspect. The cooling process is extended after the onset of helium rain, resulting in an effective temperature of about 100 K after 5 Gyr.

In Figure \ref{fig:comparisons}, we compare the evolutionary profiles at 5 Gyr for giant planets with masses of $M_p =\{0.20, 0.15, 0.10\} \, M_J$. Because $M_c$ and $M_z$ are kept constant, the total heavy element mass fractions differ for the different planets. Due to this, the size of the core also varies and the density jump lies at different locations across the core-envelope boundary (top-left panel). We start with different initial entropies for each planet while maintaining hot-start initial conditions with an initial planet radius of 2$-$3 times $R_J$. Higher-mass planets exhibit sustained elevated temperatures in their cores, while the envelope remains hotter for lower-mass planets (as depicted in the top-right panel). This is due to the higher metal content by mass for our low-mass planets. Correspondingly, central pressure increases with higher $M_p$ (as depicted in the bottom-right panel). Our simulations indicate that the core size and the extent of the immiscibility region affect the amount of helium depletion. Since lower-mass planets have narrower immiscibility regions, their $Y$ in the deeper layers is lower and their $Y_{\rm atm}$ is higher (top middle panel). We observe significant helium depletion for the $0.2 M_J$ planet, with its immiscibility region extending to the core-envelope boundary and its $Y_{\rm atm}$ reaching $\sim0.2$.  The evolution of $T_{\rm eff}$ and $T_{\rm int}$ for the different masses are shown in the bottom-left panel. The impact of helium rain in extending the cooling process is evident across all models. Nevertheless, their effective temperatures become indistinguishable in the later stages of evolution. Notably, the internal temperature is lower for the lower mass planet. However, the evolution is influenced by the atmospheric boundary condition employed, which isn't custom-tailored for these specific models, but serves as a proxy to showcase the impact of a realistic representation.

\begin{figure*}
    \centering
    \includegraphics[scale=0.445]{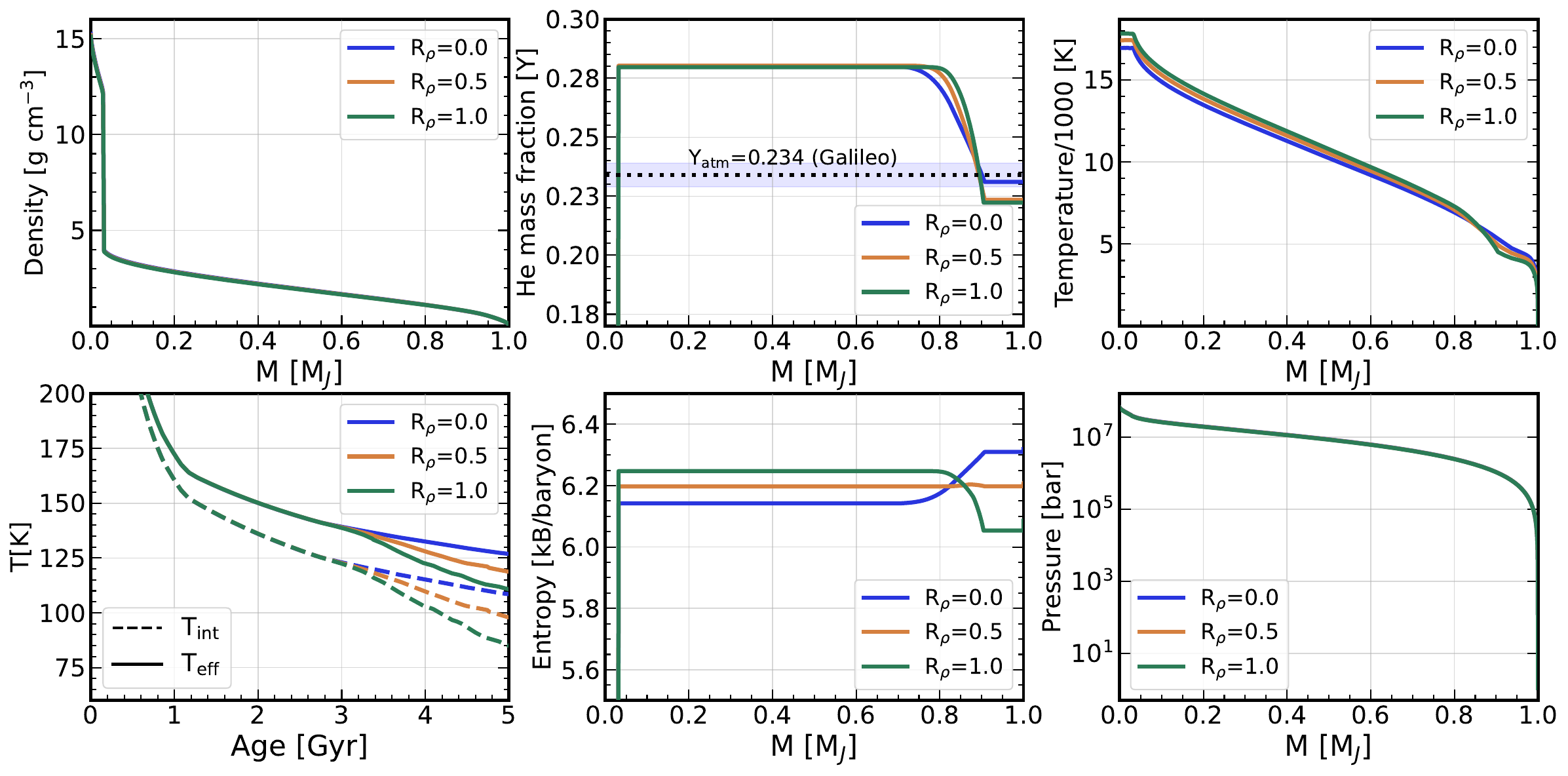}
    \caption{A comparison of the different state variables, and the effective and internal temperatures at 5 Gyr for the example $1 M_J$ model with varying $R_{\rho}$ parameter. The blue line corresponds to $R_{\rho}=0$ (Schwarzschild criterion), the green line corresponds to $R_{\rho}=1$ (Ledoux criterion), and the orange corresponds to $R_{\rho}=0.5$ (crudely, semiconvection). All other model parameters are the same as in Figure \ref{fig:1mjup}. With helium-rain, $R_{\rho} = 0$ produces a hotter planet than the rest.}
    \label{fig:R0_comparisons}
\end{figure*}
\subsection{Evolution of $1$ M$_J$ Giant Planet}
We now discuss the evolution of an artificial Jupiter-mass object. This model also consists of a core of mass $M_c=10 M_{\oplus}$ made up of the same composition described above, with $M_z=5 M_{\oplus}$ in the envelope made of water-ice. We use the CMS19+HG23 EOS for modeling the H-He envelopes and the L0911 miscibility curve without any temperature shift for modeling helium-rain using scheme B. However, we lower the $\mathcal{H}_r$ to $5\times10^7$ cm in this model to match the \textit{Galileo} constraint. We employ the $3.16\times$ solar metallicity Jupiter atmospheric boundary condition \citep{Chen2023} to calculate the luminosity lost from the surface. The model atmosphere includes ammonia clouds with a particle size of 1 micron and a redistribution parameter $P_{\rm irr}=0.5$, which redistributes between the day and night regions the intercepted solar flux originating at a distance of 5.2 A.U. 

Figure \ref{fig:1mjup} illustrates evolutionary profiles of different quantities with panels corresponding to the same variables discussed in the 0.25 M$_J$ case. The density profile is represented in the top-left panel, where the jump marks the transition from the envelope to the core, each having a different composition. The compact core roughly makes up 3\% of the planet's mass. After an initial rapid contraction phase, the planet continues to cool and the central temperature reduces from 48,000 K to 16,700 K in 5 Gyr. Concurrently, the central pressure increases to about 64 Mbar. The top-right panel displays the temperature profiles. The helium mass fraction and entropy profiles at different points in the planet's evolution are depicted in the middle top and bottom panels, respectively. Before the onset of helium rain, the planet undergoes homogeneous evolution. Around 2.4 Gyr into its evolution, the planet begins to separate into different layers, with helium-rain continuing thereafter because its $P-T$ profiles then lie below the L0911 immiscibility curve. During this period, the helium mass fraction in the atmosphere decreases, eventually reaching approximately $Y_{\rm atm}\sim0.231$ by the 5 Gyr mark. The convective stability of the planet is assessed using the Schwarzschild criterion, indicating that the planet maintains full convection throughout, which facilitates efficient heat transport across the demixing zone. This process contributes to heating the outer envelope and retarding the decrease in the surface entropy.  The result is an entropy on the periphery of $6.31$ kB/baryon,  while that in the interior envelope is at $6.1$ kB/baryon. 

The evolution of $T_{\rm eff}$ and $T_{\rm int}$ are shown in the bottom left panel. The kink observed at 1.2 Gyr is caused by the formation of ammonia clouds at an atmospheric temperature of $\sim 160$ K. This prolongs the cooling of the planet which results in $T_{\rm eff} \sim 126$ K and $T_{\rm int}\sim 107$ K by the end of our simulation. 

Next, depicted in Figure \ref{fig:R0_comparisons}, we adjust the parameter $R_{\rho}$ to study its influence on the evolutionary profiles. Here, $R_{\rho}=0$ corresponds to the Schwarzschild criterion, illustrated in blue, $R_{\rho}=1$ represents the Ledoux criterion, shown in green, and $R_{\rho}=0.5$, shown in orange, serves to explore in a very approximate fashion the effect of efficient semiconvection. While pressure, temperature, and density profiles exhibit similarities at 5 Gyr, variations in entropy, helium fraction, and $T_{\rm eff}$ are observed. These models experience helium rain, leading to subtle differences in the atmospheric helium content towards the end, which decreases with an increase in $R_{\rho}$ from 0 to 1. Specifically, the model with $R_{\rho}=0$ experiences effective heat transport through the helium-rain zone, resulting in a hotter outer envelope. This can be observed by looking at the surface entropies in the bottom right panel. In contrast, employing the pseudo-semiconvection criterion or the Ledoux criterion reduces heat transport efficiency through the helium-rain zone, causing the surface to cool more, while the interior remains hotter. The evolution of $T_{\rm eff}$ and $T_{\rm int}$ further highlights the differences in heat transport, depicted in the bottom-left panel, where a decreasing trend is observed with an increase in $R_{\rho}$, in agreement with \cite{Mankovich2016}. 

Collectively, these test evolutionary simulations provide a glimpse at the capabilities of \texttt{APPLE}.  There are many parameters that can be altered to study the broad spectrum of possibilities, and we have shown here but a few examples. The ultimate goal is to identify the crucial features of giant planets and these are naturally informed by the recent measurements of the Jovian planets of our solar system. In the near future, we will apply \texttt{APPLE} to assess the likely range of physical characteristics that determine the evolution, or range of likely evolutionary pathways, of giant exoplanets, Jupiter, and Saturn and these investigations will be informed by the knowledge culled using \textit{Galileo}, \textit{Juno}, and \textit{Cassini}. We will also be guided by theories of planet formation \citep{Bodenheimer1986,Pollack1996,Alibert2005,DAngelo2011,Modasini2012}, with one goal being to help discriminate the various birth scenarios of this important class of planet.

\section{Summary}
\label{sec:summary}
With this paper, we have introduced \texttt{APPLE}, a new planetary evolution code purpose-built to study the evolution of giant planets. This software tool is tailored specifically to address the emerging complexities of planetary evolution revealed by recent planetary probes. Key to the development of \texttt{APPLE} has been the incorporation of the most current equations of state for hydrogen, helium, ices, and rock; the inclusion ice/rock cores and metals in the gaseous envelope; the embedding of conservative prescriptions for helium rain and hydrogen/helium immiscibility; the use of detailed atmosphere boundary tables (which also provide self-consistent Bond and geometric albedos and multi-frequency spectra); and options for envelope metal gradients and stably-stratified regions. The pre-calculated atmosphere tables employ the latest molecular opacities and a sophisticated spectral and atmosphere algorithm already validated over the last twenty years of exoplanet and brown dwarf research.

{We have crafted \texttt{APPLE} to build on modern stellar evolutionary codes, with new approaches to semi-convection and helium rain, that, while not
perfect, are embedded in the code with some flexibility.  \texttt{APPLE} also
has a sophisticated approach to the atmospheric boundary condition
(see \citet{Chen2023}), that builds on the original work of Burrows et al.
(1997) that has been used by \citet{Fortney2011}; \citet{marley2021_sonora}; and \citet{Baraffe1998} since then, but also incorporates insights from Galileo atmosphere data in the context of models of Jupiter itself.}

By integrating these advanced features from the start, \texttt{APPLE} will facilitate the simulation in an integrated fashion of the complex processes of giant planet evolution that have emerged to be of central importance since the revelations of $Juno$, $Cassini$, and $Galileo$. \texttt{APPLE} is now poised to create a new generation of giant exoplanet and Jovian planet evolutionary models and, hopefully, to elevate this fascinating and important topic to the next level of physical fidelity.

\section*{Acknowledgments}
%\begin{acknowledgments}
We thank Jonathan Fortney, Ronald Redmer, and Pascale Garaud for their useful insights and help during the execution of this project. {AS thanks Romain Teyssier for valuable discussions on numerical methods}. Funding for this research was provided by the Center for Matter at Atomic Pressures (CMAP), a National Science Foundation (NSF) Physics Frontier Center, under Award PHY-2020249. Any opinions, findings, conclusions, or recommendations expressed in this material are those of the author(s) and do not necessarily reflect those of the National Science Foundation. YS is supported by a Lyman Spitzer, Jr. Postdoctoral Fellowship at Princeton University.
%\end{acknowledgments} 

\appendix
\section*{{Numerical Methods}}
\begin{figure*}
    \centering
    \includegraphics[scale=0.52]{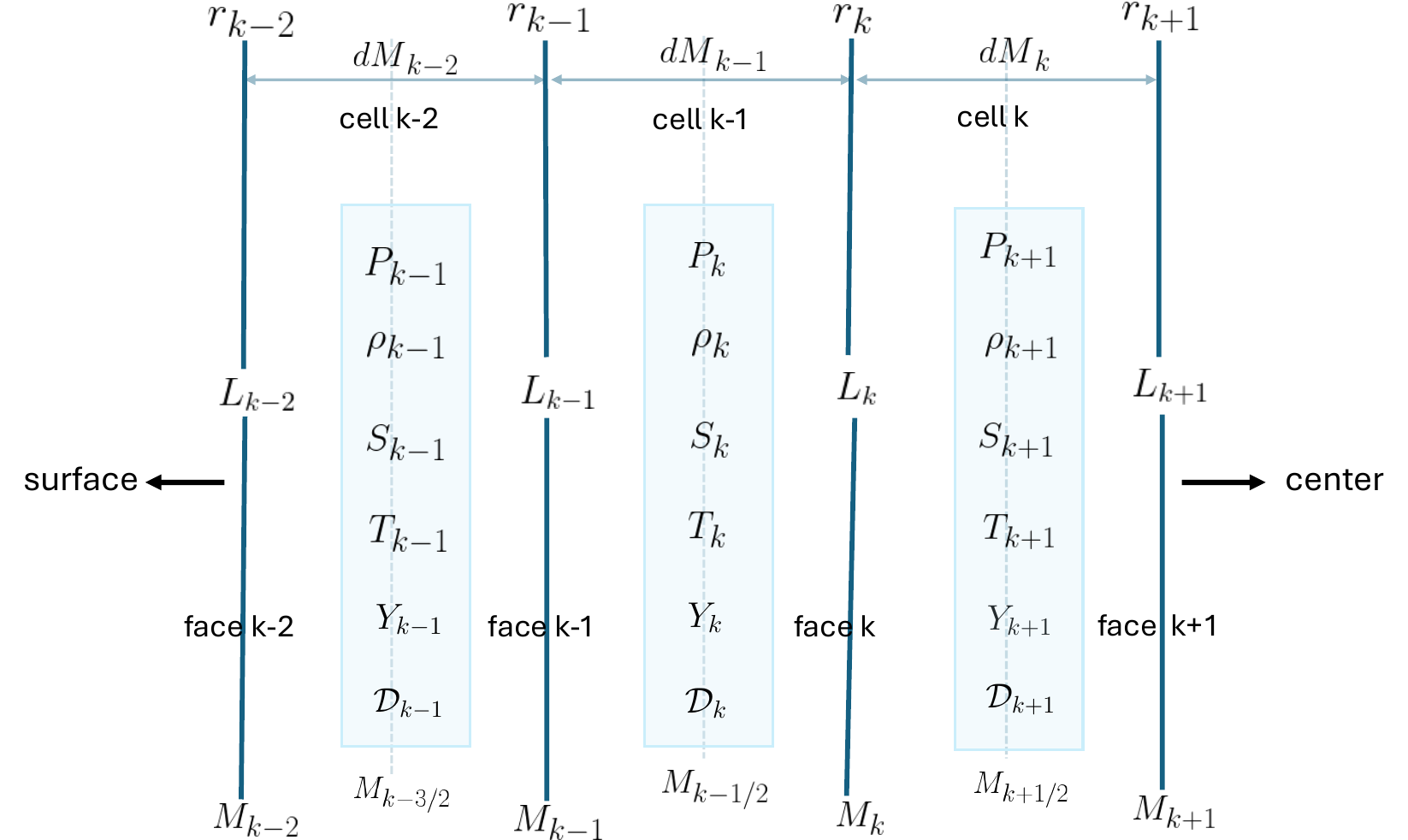}
    \caption{Finite difference grid showing locations of various thermodynamic variables and the Lagrangian coordinate $M$. The radius $r$, mass coordinate $M$, and the luminosities $L$ are evaluated at cell faces, while all other quantities are evaluated at cell centers.}
    \label{fig:grid}
\end{figure*}

Practical stellar evolution codes, such as MESA \citep{Paxton2011,Paxton2013,Paxton2015,Paxton2018,Paxton2019}, Cambrdige STARS \citep{Eggleton1971}, YREC \citep{Demarque2007}, GARSTEC \citep{Weiss2008}, and CESAM \citep{Morel2008}, and ASTEC \citep{Dalsgaard2008} use iterative methods instead of direct integration. Within the Henyey formalism \citep{Henyey1964}, mass zones ($M_r$) are set in advance and are not modified during the iterations. The convergence process becomes significantly faster when the timesteps are selected to be sufficiently small. The thermodynamic variables $P, \rho, T, S,$ and element mass fractions $X_i$ are defined at the cell centers while the masses and the radii are defined at cell faces (see Figure \ref{fig:grid}). We use a mass grid $M$\footnote{We drop the subscript $r$ from $M_r$ for convenience.}, with N+1 dimensions and follow the convention in which the indices are numbered from $k=0$ (surface) to $k=N$ (center).

\section*{Hydrostatic equilibrium}

In stellar and planetary evolution, it is often useful to employ log quantities. Let us also define $x = \ln r$, $y = \ln p$, and $q = \ln \rho$. With these definitions, we recast the hydrostatic equilibrium equations \ref{eq:1} and \ref{eq:2} to the following form:
\begin{align}
&\frac{d y}{dM} = -\frac{G M}{4\pi}\exp(-4x - y) + \frac{\Omega^2}{6\pi}\exp(-y-x)\\
&\frac{d x}{d M} = \frac{1}{4\pi}\exp(-3x - q_r) \, ,
\end{align}
where $\Omega$ is obtained by conserving angular momentum. The finite difference scheme for the above set of equations for cells $1\leq k\leq N-2$ are:
\begin{align}
& G_{k} \equiv y_{k-1} - y_{k} + \frac{G}{8\pi}(M_{k-1}-M_{k})\exp\left({-\frac{1}{2}(y_{k-1}+y_{k})-4x_{k}}\right) - \frac{(M_{k-1}-M_{k})\Omega_k^2}{12\pi}\exp\left(-x_k-\frac{1}{2}(y_{k-1}+y_{k})\right)= 0\\
& H_k \equiv x_k - x_{k+1} -\frac{1}{4\pi}(M_k - M_{k+1})\exp\left(-\frac{3}{2}(x_k + x_{k+1})-q_k\right) = 0\, .
\end{align}
At the boundaries, we treat the hydrostatic equations differently. At the surface ($k=0$),
\begin{align}
    &G_0 \equiv \exp(y_0) - P_{\rm atm} - \frac{G M_0}{8\pi}(M_0-M_1)\exp(-4x_0)=0\\
    &H_0 \equiv x_0 - x_{1} -\frac{1}{4\pi}(M_0 - M_{1})\exp\left(-\frac{3}{2}(x_0 + x_{1})-q_0\right) = 0 \, ,
\end{align}
while at the center ($k=N-1$),
\begin{align}
    & G_{N-1}\equiv \exp(y_{N-2}) - \exp(y_{N-1}) + \frac{G}{2}\left(\frac{4\pi}{3.0}\right)^{1/3}M_{N-1}^{2/3}\rho_{N-1}^{4/3} = 0 \\
    & H_{N-1}\equiv x_{N-1} - \frac{1}{3}\bigg(\ln \bigg[\frac{3M_{N-1}}{4\pi}\bigg] -q_{N-1} \bigg) = 0 \, .
\end{align}
At $k=N$, $x_N = 0$. Thus, there are 2N unknown variables, i.e. $\{(x_0,x_1,..,x_{N-1}),(y_0,y_1,...,y_{N-1})\}$, and 2N unknown equations $\{(G_0,G_1,..,G_{N-1}),(H_0,H_1,...,H_{N-1})\}$. These are solved using an iterative Newton-Raphson scheme discussed in section \S \ref{sec:methods}.

\section*{Transport}
The transport module solves equations \ref{eq:3a}, \ref{eq:3b}, and \ref{eq:4} simultaneously. These are time evolution equations, where $n$ denotes the current timestep and $n-1$ denotes the previous timestep. Generally, the coefficients and the radius in these equations are treated explicitly, while any term involving $S$, $Y$, and $Z$ is treated implicitly. The finite difference scheme for $0\leq k \leq N-1$ is given by
\begin{align}
    &E_k \equiv S^{n}_k - S^{n-1}_k + \sum_i \frac{\partial U_k}{\partial X_i}\bigg|_{S,\rho}\frac{(X_i^{n}(k)-X_i^{n-1}(k))}{T_k^{n-1}} + \frac{\Delta t}{T_k^{n-1}(M_k-M_{k-1})}\big(4\pi r_k^2\flux^n_{k} - 4\pi r^2_{k-1}\flux^n_{k-1}\big) = 0\\
    %&D_k \equiv Y^{n}_k - Y^{n-1}_k + \frac{\Delta t}{M_k-M_{k-1}}\left(4\pi r^2\rho \big|_k^{n-1}\mathcal{D}_{k+1/2}\frac{Y_{k+1}^n-Y_{k}^n}{r_{k+1/2} - r_{k-1/2}} - 4\pi r^2\rho \big|_{k-1}^{n-1}\mathcal{D}_{k-1/2}\frac{Y_k^n-Y_{k-1}^n}{r_{k-1/2} - r_{k-3/2}} + {F_D}\big|_{k}-{F_D}\big|_{k-1}\right)=0\, .
    &D_k \equiv Y^{n}_k - Y^{n-1}_k + \frac{\Delta t}{M_k-M_{k-1}}\left(4\pi r_k^2F_i^n(k) - 4\pi r_{k-1}^2 F_i^n (k-1)\right)=0\, ,
\end{align}
where $F_i^n$ is the species flux including the diffusive and advective terms. The energy fluxes $\mathcal{F}^n$ are composed of radiation, conduction, and convection and must be implicitly treated. We expand the individual components as follows:
\begin{align}
    \mathcal{F}_k^{\rm cd/r} &= -\frac{\lambda_{\rm cd/r}}{r_{k+1/2}-r_{k-1/2}}\left(T_{k+1}^n - T_{k}^n\right) \\
    &= -\frac{\lambda_{\rm cd/r}}{r_{k+1/2}-r_{k-1/2}}\left(T_{k+1}^{n-1} + \frac{\partial T}{\partial S}\bigg|_{\rho,Y}\Delta S_{k+1} + \frac{\partial T}{\partial Y}\bigg|_{\rho,S}\Delta Y_{k+1} - T_{k}^{n-1}-\frac{\partial T}{\partial S}\bigg|_{\rho,Y}\Delta S_{k}-\frac{\partial T}{\partial Y}\bigg|_{\rho,S}\Delta Y_k\right) \, ,
\end{align}
where cd/r stands for conduction/radiation (where we have dropped the $\Delta Z$ terms here and below for simplicity). For the convection flux, we have,
\begin{align}
    \mathcal{F}_{k}^{\rm conv} = \rho_{k+1/2}^{n-1} T_{k+1/2}^{n-1} \sqrt {\frac{ g H_p^4}{32C_p}} \left[{\rm max}\left(0,-\frac{S_{k+1}^{n}-S_{k}^{n}}{r_{k+1/2}-r_{k-1/2}}+\left(\frac{\partial S}{\partial Y}\bigg|_{P,T}\right)\frac{Y_{k+1}^n-Y_k^n}{r_{k+1/2}-r_{k-1/2}}\right)\right]^{3/2} \, .
\end{align}
$\mathcal{F}_{k-1}$ can be obtained by simply replacing the index $k$ by $k-1$ in the above equations. At the center, all the fluxes go to zero automatically as $r_{N} = 0$. In the case of a core, the species flux at the core-envelope boundary is set to zero. At the surface, we do not allow any $Y$ flux. However, the energy flux $\mathcal{F}_{\rm surf}$ is set by the atmosphere boundary condition and is treated implicitly,
\begin{align}
 \mathcal{F}_{\rm surf} &= \sigma T_{\rm int}^4 (n)=\sigma T_{\rm int}^4 (n-1) + 4\sigma T_{\rm int}^3 (n-1) \frac{\partial T_{\rm int}}{\partial S}\bigg|_0(S_0^{n} - S_0^{n-1}) + 4\sigma T_{\rm int}^3 (n-1) \frac{\partial T_{\rm int}}{\partial Y}\bigg|_0(Y_0^{n} - Y_0^{n-1})\, .
\end{align}
Again, there are 2N unknown equations $\{(E_0,E_1,..,E_{N-1}),(D_0,D_1,...,D_{N-1})\}$ with 2N unknown variables $\{(S_0,S_1,..,S_{N-1}),(Y_0,Y_1,...,Y_{N-1})\}$, which are solved simultaneously using the Newton-Raphson solver. The overall scheme is implicit and the transport sector update itself is solved in such a way that the flux into a zone is exactly equal to the flux out of the corresponding adjacent zone.  In this way, energy (and in fact composition) is conserved to machine accuracy during the thermal and compositional updates.  The operator splitting with the hydrostatic solve slightly breaks this (except for composition), but as shown in \S\ref{sec:methods} overall energy conservation is quite good ($<1$\%). 

%\section{Software Infrastructure}
\label{sec:numerical}

\bibliography{references}{}
\bibliographystyle{aasjournal}

\end{document}